\documentclass[twocolumn,english,pra]{revtex4-2}
\usepackage[utf8]{inputenc}
\usepackage{color}
\usepackage{babel}
\usepackage{mathrsfs}
\usepackage{amsmath}
\usepackage{amsthm}
\usepackage{amssymb}
\usepackage{stackrel}
\usepackage{graphicx}
\usepackage{esint}
\usepackage{xargs}[2008/03/08]
\usepackage[unicode=true,pdfusetitle,
 bookmarks=true,bookmarksnumbered=false,bookmarksopen=false,
 breaklinks=false,pdfborder={0 0 0},pdfborderstyle={},backref=false,colorlinks=true]
 {hyperref}

\makeatletter

\providecommand{\tabularnewline}{\\}

\usepackage[justification=raggedright]{caption}

 {
      \theoremstyle{plain}
      }

\usepackage{amsmath}
\usepackage{orcidlink}

\usepackage{ragged2e}
\DeclareCaptionJustification{justified}{\justifying}
\@ifundefined{showcaptionsetup}{}{%
 \PassOptionsToPackage{caption=false}{subfig}}
\usepackage{subfig}
\makeatother

\begin{document}
\title{Collapse and revival structure of information backflow for a central
spin coupled to a finite spin bath}
\author{Jingyi Fan\orcidlink{0000-0001-5816-1235}}
\affiliation{School of Physics, Sun Yat-sen University, Guangzhou, Guangdong 510275,
China}
\author{Shengshi Pang\orcidlink{0000-0002-6351-539X}}
\email{pangshsh@mail.sysu.edu.cn}

\affiliation{School of Physics, Sun Yat-sen University, Guangzhou, Guangdong 510275,
China}
\begin{abstract}
The Markovianity of quantum dynamics is an important property of open
quantum systems determined by various ingredients of the system and
bath. Apart from the system-bath interaction, the initial state of
the bath, etc., the dimension of the bath plays a critical role in
determining the Markovianity of quantum dynamics, as a strict decay
of the bath correlations requires an infinite dimension for the bath.
In this work, we investigate the role of finite bath dimension in
the Markovianity of quantum dynamics by considering a simple but nontrivial
model in which a central spin is isotropically coupled to a finite
number of bath spins, and show how the dynamics of the central spin
transits from non-Markovian to Markovian as the number of the bath
spins increases. The non-Markovianity is characterized by the information
backflow from the bath to the system in terms of the trace distance
of the system states. We derive the time evolution of the trace distance
analytically, and find periodic collapse-revival patterns in the information
flow. The mechanism underlying this phenomenon is investigated in
detail, and it shows that the period of the collapse-revival pattern
is determined by the competition between the number of the bath spins,
the system-bath coupling strength and the frequency detuning. When
the number of bath spins is sufficiently large, the period of the
collapse-revival structure as well as the respective collapse and
revival times increase in proportion to the number of the bath spins,
which characterizes how the information backflow decays with a large
dimension of the bath. We also analyze the effect of the system-bath
interaction strength and frequency detuning on the collapse-revival
patterns of the information flow, and obtain the condition for the
existence of the collapse-revival structure. The results are illustrated
by numerical computation.
\end{abstract}
\maketitle
\newcommandx\tdt[2][usedefault, addprefix=\global, 1=\rho_{1}, 2=\rho_{2}]{D_{ }\left(#1,#2\right)}%
 \newcommandx\tdsq[2][usedefault, addprefix=\global, 1=\rho_{1}(0), 2=\rho_{2}(0)]{D_{ }^{2}\left(#1,#2\right)}%
 \newcommandx\td[2][usedefault, addprefix=\global, 1=z, 2=t]{D_{#1 }\left(#2\right)}%
\newcommandx\tds[2][usedefault, addprefix=\global, 1=x, 2=t]{D_{#1 }^{2}\left(#2\right)}%
\newcommandx\tr[1][usedefault, addprefix=\global, 1=]{\mathrm{Tr}_{ #1 }}%

\section{Introduction}

A quantum system inevitably interacts with the environment in practical
applications, leading to irreversible processes and effects \citep{alicki1987}
such as the loss of quantum coherence, the dissipation of information,
the degradation of the entanglement, etc \citep{breuer2016}. Therefore,
it is important to consider the impact of the environment on the system
when we investigate the evolution of a quantum system in reality.
The evolution of a quantum system interacting with an environment
can be formally obtained by tracing out the degrees of freedom of
the environment from the joint evolution of the system and the environment,
but the derivation is usually difficult since the interaction with
the environment can be complex and the environment may have memory
effects. If the interaction between the system and environment is
sufficiently weak and the dimension of the environment is large, the
well-known Born-Markov approximation can be employed, with which the
evolution of the system becomes a Markovian process, and a neat master
equation with a Lindblad structure can be derived \citep{gorini1976,lindblad1976}.

An interesting feature of Markovian processes is that the environment
is memoryless since the correlation time of the environment is short
compared to the decoherence time of the system. As a consequence,
the information of the system lost into the environment will vanish,
and cannot flow back to the system. On the contrary, in the presence
of structured or finite environment, or strong coupling between the
system and environment, the Born-Markov approximation may fail and
the evolution of the system turns to be non-Markovian. In non-Markovian
quantum processes, the environment can have a memory effect, and backflow
of system information from the environment to the system may appear
at some time points, implying the quantum states of the system in
the past can contribute to the evolution of the system \citep{breuer2007}.

From a mathematical point of view, quantum dynamics can generally
be described by a completely positive and trace-preserving (CPTP)
map. For a Markovian quantum process, a CPTP map can be decomposed
into the product of consecutive CPTP maps of arbitrary division of
the evolution time, which implies that the Markovian dynamical maps
form a semigroup. In contrast, the CPTP divisibility does not hold
for non-Markovian quantum processes. However, it is highly nontrivial
to determine whether a quantum process is Markovian or non-Markovian
by investigating its CPTP divisibility. Some reliable ways have been
proposed to witness and quantify the non-Markovianity of quantum dynamics,
based on the monotonicity of specific physical quantities under CPTP
maps. One such idea originates from the fact that the trace distance
between any two quantum states, which characterizes the information
that the quantum system carries, does not increase in a Markovian
quantum process, indicating one-way information flow from the system
to the environment. If the trace distance increases at some time points
in a quantum process, it suggests that the CPTP divisibility breaks
and the quantum process is non-Markovian, and the memory effect of
the environment occurs as there is information flowing from the environment
back into the system, therefore the information backflow can serve
as a measure of non-Markovianity \citep{laine2010,breuer2009,breuer2012}.
Another idea is rooted in the fact that the entanglement between the
system and an ancilla never increases in a Markovian process on the
system, so if one observes an increase in the entanglement between
the system and an ancilla, it immediately tells that the process is
non-Markovian, and the increase of entanglement can quantify the degree
of non-Markovianity of the quantum process \citep{rivas2010,rivas2014}.
There are other useful non-Markovianity measures based on different
monotonic physical quantities, such as relative entropy of coherence
\citep{wu2020,vedral2002,he2017}, fidelity \citep{rajagopal2010,vasile2011},
Fisher information flow \citep{lu2010,song2015}, etc.

A critical assumption for the Born-Markov approximation of open systems
is that the environment has infinite or sufficiently large degrees
of freedom \citep{breuer2007} so that the time correlation of the
bath can decay strictly. Thus, an interesting question arises: How
does the non-Markovianity of open quantum systems change with the
dimension of the environment if the environment has finite dimension,
and how does non-Markovian quantum dynamics transit to Markovian quantum
dynamics when the dimension of the environment goes to infinity?

The purpose of this work is to investigate the above problem by considering
the non--Markovianity of a central spin coupled to a finite number
of bath spins. We study the effect of the number of bath spins (thus
the bath dimension) on the non-Markovianity of the central spin dynamics,
as well as the effects of the coupling strength, the detuning, the
environment temperature, etc., and show how the non-Markovianity of
the central spin dynamics changes with the number of bath spins. To
simplify the problem, we are mainly interested in the case that the
couplings between the central qubit and all the bath qubits are all
identical and the initial state of the bath is symmetric between all
the bath spins so that the bath state bears high symmetry and can
always be spanned by the Dicke states of the bath spins. It is noteworthy
that the dimension of the bath is not the only ingredient that determines
the Markovianity of quantum dynamics and there exist quantum systems
that exhibit non-Markovian behavior even when the bath dimension is
infinite. But as we are mainly interested in the transition of quantum
dynamics from non-Markovian to Markovian when the bath dimension goes
from finite to infinite in this work, we will focus on the cases where
the system dynamics is Markovian when the bath is infinite dimensional,
and it will be shown later that the dynamics of the central spin in
the current model is indeed Markovian when there are infinite bath
spins simultaneously coupled to the central spin.

There are different measures for the non-Markovianity of quantum dynamics
as reviewed above. In this paper, we take the information backflow
to quantify the degree of non-Markovianity \citep{breuer2009}, which
is characterized by the increase of the trace distance between two
states of the quantum system in the open system dynamics. \textcolor{black}{While
the information backflow can occur in the current finite-dimensional
bath model similar to that in other infinite-dimensional bath models,
the results of this work reveal properties of the system dynamics.}
In particular, we show interesting collapse-revival patterns in the
information flow when the number of bath spins is finite, in analogy
to the atomic population inversion in the Jaynes-Cummings model for
a single-mode photonic field \citep{eberly1980,narozhny1981}, and
that it occurs periodically over the system evolution time, indicating
a non-vanishing oscillation in the information flow between the system
and the bath. To characterize the collapse-revival phenomenon, we
analytically obtain the envelopes of the oscillations of the information
flow for arbitrary initial states of the central qubit, and derive
the periods and amplitudes of the collapse-revival pattern in general.
We find the relation between the periods (and amplitudes) of the collapse-revival
patterns and the number of bath qubits, and show the effects of interaction
strength, frequency detuning and bath temperature, on the system dynamics
as well, which leads to an existence condition for the collapse-revival
patterns of information flow. Finally, we analyze how the transition
from non-Markovian dynamics to Markovian dynamics occurs when the
number of bath qubits goes to infinity.

The paper is organized as follows. In Sec.\,\ref{sec:Preliminaries},
we give preliminaries for the evolution of open quantum systems and
the measure of non-Markovianity. In Sec.\,\ref{sec:Central-spin-coupled},
we introduce the isotropic central spin model and derive the reduced
dynamics of the central spin. Section \ref{sec:DYNAMICS-OF-NON-MARKOVIANITY}
is devoted to obtaining the trace distance between two states of the
central spin and exhibiting the collapse-revival patterns of the information
flow for different initial states of the system. Detailed analysis
of typical time scales such as the periods, the collapse time and
the revival time of the information flow are given in Sec.\,\ref{sec:Characteristic-time-scales},
and the dependence of the non-Markovianity of the central spin on
the number of bath qubits as well as the system-bath interaction and
frequency detuning are discussed. Finally the paper is concluded in
Sec.\,\ref{sec:Conclusion-and-outlook}.

\section{Preliminaries\label{sec:Preliminaries}}

In this section, we briefly introduce some fundamental concepts of
open quantum system theory relevant to the current research.

\subsection{General dynamics of open quantum systems}

In open quantum systems, the system inevitably interacts with an external
bath. The total Hamiltonian of the system and the bath can be written
as
\begin{equation}
H_{tot}=H_{s}+H_{b}+H_{sb},
\end{equation}
where $H_{s}$ and $H_{b}$ are the Hamiltonians of the system and
the bath respectively, and $H_{sb}$ is the interaction Hamiltonian
that describes the coupling between the system and the bath. The most
general interaction Hamiltonian $H_{sb}$ can be decomposed into a
sum of the products of system and bath operators
\begin{equation}
H_{sb}=\sum_{k}S_{k}\otimes B_{k},\label{Hi1}
\end{equation}
where $S_{k}$ and $B_{k}$ are the system and bath operators respectively.
Such a decomposition of the interaction Hamiltonian is always possible,
and the operators $S_{k}$ and $B_{k}$ can always be chosen to be
Hermitian due to the Hermiticity of the interaction Hamiltonian.

Suppose the initial state is factorized between the system and the
bath $\rho_{sb}\left(0\right)=\rho_{s}(0)\otimes\rho_{b}(0)$. The
joint evolution of system and bath after time $t$ can be written
as

\begin{equation}
\rho_{sb}\left(t\right)=U\left(t\right)\rho_{s}(0)\otimes\rho_{b}(0)U^{\dagger}\left(t\right),
\end{equation}
where $U\left(t\right)=e^{-iH_{tot}t}$ is the unitary dynamical evolution
operator under the total Hamiltonian. The reduced density matrix of
the system at time $t$ can be derived by tracing out the bath from
the joint density matrix $\rho_{sb}\left(t\right)$, and the reduced
evolution of the system can be written as

\begin{equation}
\rho_{s}\left(t\right)=\mathrm{Tr}_{b}\left[\rho_{sb}\left(t\right)\right]=\Lambda\left(t,0\right)\rho_{s}\left(0\right),
\end{equation}
where $\Lambda\left(t,0\right)$ is a CPTP dynamical map which can
be described by the Kraus operator sum representation,
\begin{equation}
\begin{aligned}\Lambda\left(t,0\right)\rho_{s}(0)= & \sum_{i,j}K_{i,j}(t)\rho_{s}(0)K_{i,j}^{\dagger}(t),\\
K_{i,j}(t)= & \sqrt{\lambda_{i}}\langle e_{j}|U(t)|\lambda_{i}\rangle,
\end{aligned}
\end{equation}
where $\lambda_{i}$'s and $|\lambda_{i}\rangle$'s are the eigenvalues
and eigenstates of the initial density matrix of the bath $\rho_{b}(0)$,
and $|e_{j}\rangle$'s are a set of arbitrary orthogonal basis states
of the bath.

\subsection{Non-Markovianity of quantum dynamics}

The concept of Markovianity \citep{lindblad1976} is based on the
divisibility of the CPTP map $\Lambda\left(t,0\right)$, i.e., for
Markovian quantum dynamics, the map $\Lambda\left(t,0\right)$ can
always be written in a divisible form as 
\begin{equation}
\Lambda\left(t,0\right)=\Lambda\left(t,t_{n}\right)\Lambda\left(t_{n},t_{n-1}\right)\ldots\Lambda\left(t_{2},t_{1}\right)\Lambda\left(t_{1},0\right),\label{divisible}
\end{equation}
where the time points $0\leq t_{1}\leq\ldots\leq t_{n}\leq t$ are
arbitrary and each $\Lambda\left(t_{k+1},t_{k}\right)$, $k=0,\cdots,n-1$,
is also a CPTP map. Moreover, such a process described by a divisible
CPTP map can always be derived from some Lindblad master equation
\citep{rivas20122} 
\begin{equation}
\dot{\rho}(t)=\mathcal{L}(t)\rho(t),
\end{equation}
where the time dependent generator $\mathcal{L}(t)$ can be written
in the form
\begin{equation}
\begin{aligned}\mathcal{L}(t)\rho(t)= & -i\left[H(t),\rho(t)\right]\\
 & +\sum_{k}\gamma_{k}(t)\left(V_{k}\rho V_{k}^{\dagger}-\frac{1}{2}\left\{ V_{k}^{\dagger}V_{k},\rho\right\} \right),
\end{aligned}
\end{equation}
and $\gamma_{k}(t)\geq0$ for all time $t$. The derivation of this
master equation requires some approximations, such as the rotating
wave approximation and the Born-Markov approximation \citep{breuer2007}.

If some quantum process is not CPTP divisible, non-Markovianity emerges,
where the bath can have memory effects and the evolution of the quantum
system depends on its evolution history, not only its immediate precedent
state. How to characterize and measure the non-Markovianity of quantum
dynamics is still an interesting question in the open system theory
\citep{rivas2014,breuer2016}.

\subsection{Information backflow and non-Markovianity}

After introducing the concept of quantum non-Markovianity, it is important
to quantify the degree of non-Markovianity of quantum dynamics. A
useful approach to quantifying the non-Markovianity of a quantum process
is based on the trace distance between two states of the quantum system. 

The trace distance is a measure for the difference between two quantum
states and defined as 
\begin{equation}
\tdt=\frac{1}{2}\mathrm{Tr}\left|\rho_{1}-\rho_{2}\right|,\label{td}
\end{equation}
where $\mathrm{Tr}\left|\cdotp\right|$ denotes the trace norm and
is defined as $\left|A\right|=\sqrt{A^{\dagger}A}$. It is straightforward
to verify that $\tdt=1$ if and only if $\rho_{1}$ and $\rho_{2}$
are orthogonal while $\tdt=0$ if and only if $\rho_{1}$ and $\rho_{2}$
are completely identical.

The way that trace distance quantifies the non-Markovianity of quantum
dynamics relies on the fact that the trace distance is contractive
under CPTP quantum processes \citep{nielsen2010}, i.e., 
\begin{equation}
\tdt[\Lambda\rho_{1}][\Lambda\rho_{2}]\leq\tdt.
\end{equation}
Hence, the trace distance can never increase in a Markovian quantum
process, and any increase of the trace distance in a quantum process
immediately suggests the non-Markovianity of the process (but the
reverse is not true).

Note that the trace distance can be interpreted as a measure for the
distinguishability of the two states \citep{barnett2009,bae2015},
since the minimum error probability \citep{helstrom1976} to distinguish
two arbitrary quantum states $\rho_{1},\rho_{2}$ is given by
\begin{equation}
p_{e}=\frac{1}{2}\left(1-\mathrm{Tr}\left|\rho_{1}-\rho_{2}\right|\right).
\end{equation}
This gives the trace distance an informatics sense, and thus the change
of distinguishability of two states in a quantum process can be interpreted
as the gain and loss in the information of the system \citep{breuer2009}.

In detail, a decrease in the trace distance between two states indicates
a decrease in the distinguishability of the system, implying the information
of the system flows to the bath, while an increase in the trace distance
indicates a backflow of the information from the bath to the system.
Based on the contractivity of CPTP maps and the CPTP divisibility
of quantum Markovian process, the information can flow only from the
system to the bath in a Markovian quantum process, and if one observes
any backflow of information from the bath to the system, he or she
can immediately tell that the CPTP divisibility breaks and the quantum
process is non-Markovian.

The change rate of the trace distance at time $t$ associated with
a pair of initial states $\rho_{1}\left(0\right)$ and $\rho_{2}\left(0\right)$
can be defined by

\begin{equation}
\sigma_{\rho_{1,2}\left(0\right)}(t)=\frac{dD\left(t\right)}{dt},\label{rate of if}
\end{equation}
where $D\left(t\right)$ denotes trace distance at time $t$ given
two arbitrary initial states $\rho_{1}\left(0\right)$ and $\rho_{2}\left(0\right)$.
Equation \eqref{rate of if} can be interpreted as the rate of information
flow, and a negative rate $\sigma_{\rho_{1,2}\left(0\right)}(t)<0$
indicates information flow from the system to the bath while a positive
rate $\sigma_{\rho_{1,2}\left(0\right)}(t)>0$ indicates information
backflow from the bath to the system.

A non-Markovianity measure based on the total growth of the trace
distance over the whole evolution is proposed in Ref. \citep{breuer2009},
which is defined as
\begin{equation}
\begin{aligned}\mathfrak{\mathscr{\mathcal{N}}}= & \max_{\rho_{1,2}\left(0\right)}\int_{\sigma>0}dt\sigma\left(t,\rho_{1,2}\left(0\right)\right)\\
= & \max_{\rho_{1,2}\left(0\right)}\sum_{i}\left[D\left(b_{i}\right)-D\left(a_{i}\right)\right],
\end{aligned}
\label{measure of NM}
\end{equation}
where the integration is taken over all time intervals $\left(a_{i},b_{i}\right)$
during which $\sigma>0$ and maximized over all possible pairs of
initial states of the system. According to this definition, $\mathfrak{\mathscr{\mathcal{N}}}$
is always non-negative and could be positive if a quantum process
violates the CPTP divisibility property; therefore, a positive $\mathfrak{\mathscr{\mathcal{N}}}$
indicates and measures the non-Markovianity of a quantum process.

\section{Quantum dynamics of central spin coupled to spin bath\label{sec:Central-spin-coupled}}

In this section, we introduce the system-bath model considered in
this work and derive the reduced dynamics of the system\textcolor{black}{.}

\textcolor{black}{A variety of bath models have been proposed to describe
the environment in the open system theory, which typically includes
two main categories, a set of harmonic oscillators or a set of spins
\citep{prokofev2000}. The harmonic oscillator models were derived
from the theory of radiation \citep{fermi1932} and have been widely
used in quantum optics and condensed matter physics. Two of the most
prominent oscillator models are the spin-boson model \citep{leggett1987}
and the Caldeira-Leggett model \citep{caldeira1983}, both originating
from Feynman and Vernon's influence functional technique \citep{feynman1963}.
The former considers a two-level system interacting with a bath of
bosons as oscillators, and the latter involves a tunneling system
linearly coupled to an environment of harmonic oscillators in the
spatial or momentary degrees of freedom. So far, the dynamics of oscillator
models has been widely studied for various physical phenomena }\citep{nesi2007,fiorelli2020,watanabe2017,finney1994,divincenzo2005,weiss2012,clos2012,ferialdi2017,hartmann2019,wenderoth2021,dittrich2022,goletz2010}\textcolor{black}{.}

On the other hand, the environment consisting of multiple spins, often
termed the spin bath, received early attention in the problems of
$1/f$ noise \citep{rose1998}, Landau-Zener dynamics \citep{shimshoni1991},
the quantum tunneling of magnetization \citep{prokofev1993,tomsovic1998},
etc. It is usually applicable at low temperature,\textcolor{red}{{}
}\textcolor{black}{since an ensemble of spins with finite Hilbert
spaces is suitable to describe low energy environment }and the dynamics
is dominated by localized modes \citep{prokofev2000}. So far, most
studies on spin baths have been focused on systems of few central
spins coupled to bath spins, and the bath spins can be mapped to an
oscillator model in the weak coupling limit \citep{feynman1963,caldeira1993,tomsovic1998,prokofev2000}
and solved approximately by tracing out the bath. However, in some
scenarios such as strong interaction between the system and bath,
the weak coupling limit breaks and the interaction results in a considerably
large dimension of the joint Hilbert space for the quantum system
and the bath and the problem becomes more challenging to solve. One
typical spin bath model is the spin star model \citep{breuer2004,hutton2004,hamdouni2006,yuan2007,bortz2010,dooley2013,faribault2013,Hsieh2018,Hsieh2018a},
in which the bath spins are not interacting and the interaction only
occurs between the central spin and the bath spins, thus it is exactly
solvable due to its high symmetry. Another typical category of spin
bath models involves interacting bath spins, such as one-dimensional
arrays of spins with nearest-neighbor interactions, usually known
as the spin chain model \citep{wu2014,rossini2007,luo2011,giorgi2012,lai2008,heyl2014,lu2020}
and the Lipkin--Meshkov--Glick model \citep{lipkin1965,quan2007,yuan2007,han2020}.

In this work, we are mainly interested in a central spin model where
the central spin is coupled isometrically to a bath of identical spins
and the bath is in thermal equilibrium. We assume the initial state
of the bath to be symmetric among all the bath spins to simplify the
problem. Such an assumption will also make our results suitable for
indistinguishable bath spins.

\subsection{Hamiltonian}

\begin{figure}[h]
\includegraphics[width=8cm]{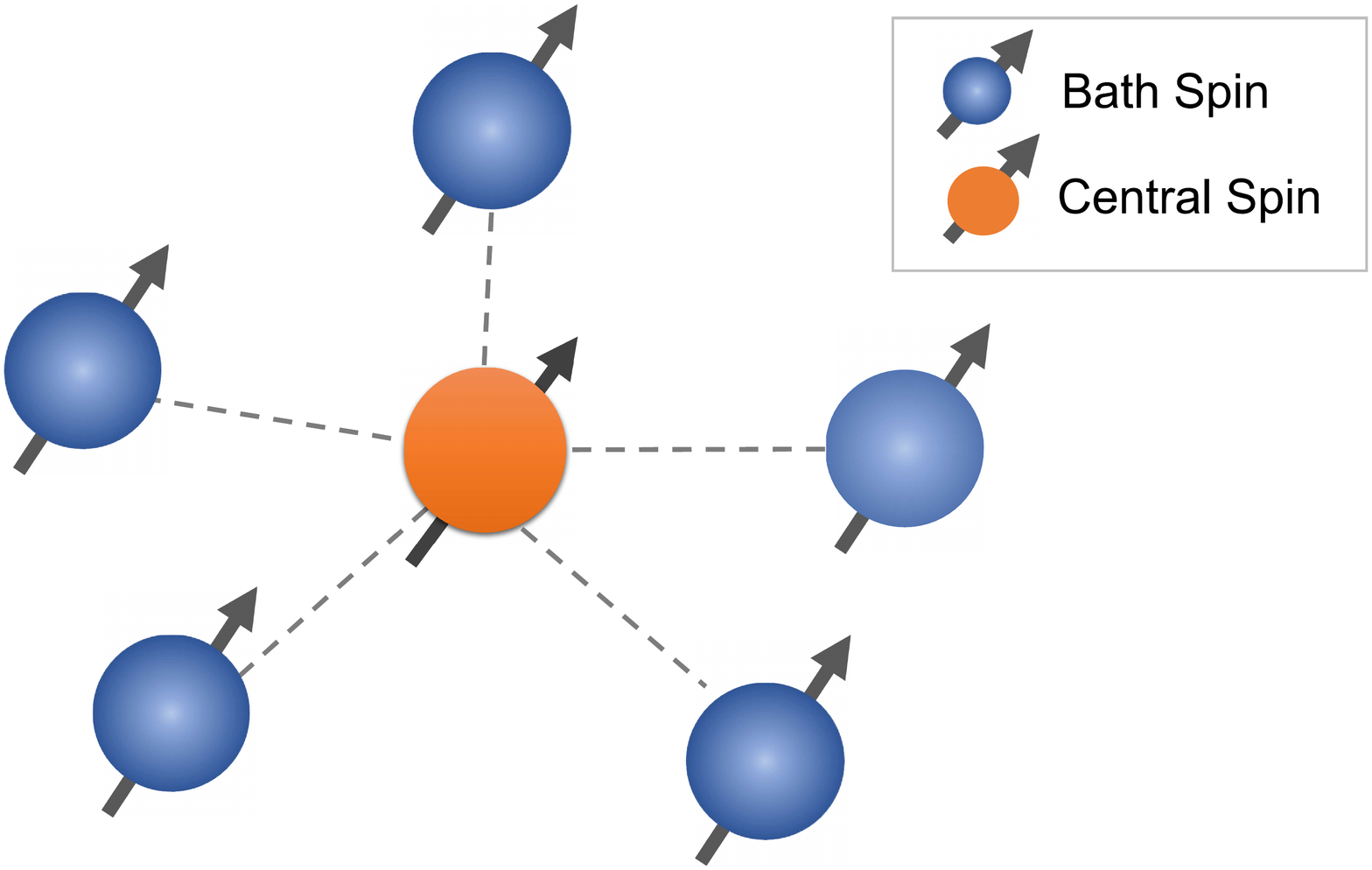} 

\caption{A sketch of the system-bath model is considered in this work. A central
spin is coupled to each spin of the bath, and there is no internal
interaction between the bath qubits. In the figure, we only show five
bath spins, while in the main text the number of spins, $N$, can
be arbitrary. The central spin is marked in red and put in the center,
and the bath spins are marked in blue and placed around the central
spin.}
\label{fig:sys-bath}
\end{figure}

Consider a composite system consisting of a central spin and a spin
bath of $N$ identical qubits. The structure of the system and the
bath is plotted in Fig.\,\ref{fig:sys-bath}, where all bath spins
only interact with the central spin, known as a star network of spins
\citep{hutton2004}.

Such a central spin and a spin bath can be described by the Hamiltonians

\begin{equation}
H_{s}=\frac{\omega_{s}}{2}\sigma_{z}^{(s)},\;H_{b}=\frac{\omega_{b}}{2}\sum_{k=1}^{N}\sigma_{z_{k}}^{(b)}.
\end{equation}
The central spin is assumed to be isotropically coupled to all components
of all bath spins with the same coupling strength, thus the interaction
Hamiltonian reads
\begin{align}
H_{sb}= & g\sum_{k=1}^{N}\left(\sigma_{x}^{(s)}\otimes\sigma_{x_{k}}^{(b)}+\sigma_{y}^{(s)}\otimes\sigma_{y_{k}}^{(b)}+\sigma_{z}^{(s)}\otimes\sigma_{z_{k}}^{(b)}\right).
\end{align}
Here $\sigma_{\alpha}^{(s)}$ and $\sigma_{\alpha_{k}}^{(b)}$ ~$(\alpha,\alpha_{k}=x,y,z)$
are the Pauli operators for the central spin and the bath spins, respectively.
The parameter $g$ denotes the coupling strength between the central
spin and the bath spins, and $\omega_{s}$, $\omega_{b}$ are the
frequencies of the central spin and bath spins, respectively.

For simplicity, we assume the initial state of the bath is symmetric
among all bath spins. Since the coupling between the central spin
and the bath spins are all identical, the bath state will always remai\textcolor{black}{n
symmetric during the evolution and can be represented by Dicke states
\citep{dicke1954}, and }the Hamiltonians of the bath and of the system-bath
interaction can be written in terms of the raising and lowering operators
of Dicke states. The Dicke states of the bath can be denoted as $\lvert J,M\rangle$
with $J=N/2$ and $M=-J,-J+1,\ldots,J-1,J$, when $J+M$ bath spins
are in the upper level and $J-M$ in the lower level. The Dicke representation
can significantly reduce the dimension of the Hilbert space from $2^{N}$
to $N+1$, thus for a finite number of bath spins, obtaining the eigenvalues
of the total Hamiltonian and studying the dynamics of the system becomes
possible. Below we use the collective spin operators $S_{\alpha}=\frac{1}{2}\sum_{i}\sigma_{\alpha}^{i}$~$(\alpha=x,y,z)$
to describe the Hamiltonian in the Dicke representation,
\begin{equation}
H_{tot}=\frac{\omega_{s}}{2}\sigma_{z}^{(s)}+\omega_{b}S_{z}+2g\left(\sigma_{+}^{(s)}S_{-}+\sigma_{-}^{(s)}S_{+}+\sigma_{z}^{(s)}S_{z}\right),\label{eq:htot}
\end{equation}
where $\sigma_{\pm}=\sigma_{x}\pm i\sigma_{y}$ and $S_{\pm}=S_{x}\pm iS_{y}$
are the raising and lowering operators of the central and bath spins,
respectively. For the Dicke representation of the total Hamiltonian,
there are invariant subspaces spanned by the pairs of states $\left\{ \lvert0\rangle\otimes\lvert J,M\rangle,\lvert1\rangle\otimes\lvert J,M-1\rangle\right\} $
with $-J+1\leq M\leq J$, which can be verified by the action of the
raising and lowing operators $S_{\pm}$ on the Dicke states,
\begin{equation}
\begin{aligned}S_{+}\lvert J,M-1\rangle & =\sqrt{\left(J-M+1\right)\left(J+M\right)}\lvert J,M\rangle,\\
S_{-}\lvert J,M\rangle & =\sqrt{\left(J-M+1\right)\left(J+M\right)}\lvert J,M-1\rangle.
\end{aligned}
\label{ladder}
\end{equation}
Then the eigenstates of $H_{tot}$ in each invariant subspace must
be the superposition of the two basis states of the invariant subspace.

One can find the reduced Hamiltonian
\begin{equation}
H_{M}=\left[\begin{array}{cc}
\frac{\omega_{s}}{2}-M\left(\omega_{b}+2g\right) & 2g\eta_{M}\\
2g\eta_{M} & -\frac{\omega_{s}}{2}-(M-1)\left(\omega_{b}-2g\right)
\end{array}\right]
\end{equation}
in the invariant subspace, where
\begin{equation}
\eta_{M}=\sqrt{\left(J-M+1\right)\left(J+M\right)},
\end{equation}
and two eigenvalues of $H_{M}$ in the invariant subspace can be obtained.
For simplicity, we denote each eigenvalue as 
\begin{equation}
\lambda_{M,\pm}=-g+\frac{2M-1}{2}\omega_{b}\pm F_{M},
\end{equation}
where $F_{M}$ is a function dependent on $M$:
\begin{equation}
F_{M}=\sqrt{G_{M}^{2}+4\eta_{M}^{2}g^{2}},
\end{equation}
and
\begin{equation}
G_{M}=\left(2M-1\right)g+\frac{\Delta}{2}.
\end{equation}
Here $\Delta=\omega_{s}-\omega_{b}$ is the frequency detuning. The
eigenstates of $H_{M}$ can also be obtained for each $-J<M\leq J+1$,
\begin{equation}
\lvert\Phi\rangle_{M,\pm}=d_{M,\pm}\lvert1\rangle\lvert J,M-1\rangle\pm{\rm sgn}(g)d_{M,\mp}\lvert0\rangle\lvert J,M\rangle,\label{eigenstate}
\end{equation}
associated with the eigenvalues $\lambda_{M,\pm}$ respectively, where
\begin{equation}
d_{M,\pm}=\sqrt{\frac{1}{2}\left(1\pm\frac{G_{M}}{F_{M}}\right)}.
\end{equation}

There are two additional eigenstates, $\lvert\Phi\rangle_{-J}=\lvert0\rangle\otimes\lvert J,-J\rangle$
and $\lvert\Phi\rangle_{J+1}=\lvert1\rangle\otimes\lvert J,J\rangle$,
and the corresponding eigenvalues are
\begin{equation}
\lambda_{-J}=(2g-\omega_{b})J-\frac{1}{2}\omega_{s},\;\lambda_{J+1}=(2g+\omega_{b})J+\frac{1}{2}\omega_{s}.
\end{equation}
Therefore, there are $4J+2$ (or equivalently $2N+2$) eigenstates
in total, in accordance with the dimension of the joint Hilbert space
of the system and bath.

\subsection{Exact time evolution}

Suppose the initial state can be factorized as the product of the
mixed states of the system and bath

\begin{equation}
\rho_{sb}\left(0\right)=\rho_{s}\left(0\right)\otimes\rho_{b}\left(0\right),
\end{equation}
and the density matrix of the central spin can be written as
\begin{equation}
\rho_{s}\left(0\right)=\frac{1}{2}\left[I+\boldsymbol{v}(0)\cdot\boldsymbol{\sigma}\right],\;\boldsymbol{v}(0)=[x(0),y(0),z(0)],
\end{equation}
where $\boldsymbol{v}(0)$ is the Bloch vector and $\boldsymbol{\sigma}=[\sigma_{x},\sigma_{y},\sigma_{z}]$
is the vector of the Pauli matrices. The bath spins are assumed to
be in a thermal equilibrium state initially and its density matrix
can be described by the Dicke states,

\begin{align}
\rho_{b}\left(0\right) & =\sum_{M=-J}^{J}\frac{\exp\left(-\frac{M\omega_{b}}{k_{\text{B}}T}\right)}{Q}\lvert J,M\rangle\langle J,M\rvert,
\end{align}
where $Q$ is the partition function,
\begin{equation}
Q=\frac{\exp\left[\frac{(J+1)\omega_{b}}{k_{\text{B}}T}\right]-\exp\left(-\frac{J\omega_{b}}{k_{\text{B}}T}\right)}{\exp\left(\frac{\omega_{b}}{k_{\text{B}}T}\right)-1}.\label{eq:partition}
\end{equation}
Throughout this paper, we assume the Boltzmann constant $k_{\text{B}}=1$.

Then the reduced density matrix of the system is
\begin{equation}
\rho_{s}\left(t\right)=\mathrm{Tr}_{b}\left[U\left(t\right)\rho_{sb}\left(0\right)U^{\dagger}\left(t\right)\right].
\end{equation}
The final density matrix of the central spin can also be represented
by a Bloch vector $\boldsymbol{v}(t)=[x(t),y(t),z(t)]$, and $\boldsymbol{v}(t)$
can be obtained as (see Appendix~\ref{sec:EVOLUTION-OF-THE})

\begin{align}
x(t) & =x_{0}X_{1}\left(t\right)+y_{0}X_{2}\left(t\right),\nonumber \\
y(t) & =y_{0}X_{1}\left(t\right)-x_{0}X_{2}\left(t\right),\label{reducedsys}\\
z(t) & =z_{0}Z_{1}\left(t\right)+Z_{2}\left(t\right).\nonumber 
\end{align}
Here $X_{1}\left(t\right)$, $X_{2}\left(t\right)$, $Z_{1}\left(t\right)$
and $Z_{2}\left(t\right)$ are functions of time $t$. As these functions
are quite lengthy, we leave the detail of these functions to Appendix
{[}see Eq.\,\eqref{eq:z2t} and \eqref{eq:ambmcm}{]}.

It may also be helpful to have a master equation for the dynamics
of the central spin. Using the formalism in Refs.~\citep{andersson2007,bhattacharya2021},
the exact master equation for the central spin is given by
\begin{align}
\dot{\rho}_{s}\left(t\right)= & i\Omega(t)[\rho_{s}\left(t\right),\sigma_{z}]+\Gamma_{d}(t)[\sigma_{z}\rho_{s}\left(t\right)\sigma_{z}-\rho_{s}\left(t\right)]\nonumber \\
 & +\Gamma_{-}(t)[\sigma_{-}\rho_{s}\left(t\right)\sigma_{+}-\frac{1}{2}\{\sigma_{+}\sigma_{-},\rho_{s}\left(t\right)\}]\label{ME}\\
 & +\Gamma_{+}(t)[\sigma_{+}\rho_{s}\left(t\right)\sigma_{-}-\frac{1}{2}\{\sigma_{-}\sigma_{+},\rho_{s}\left(t\right)\}],\nonumber 
\end{align}
with 
\begin{align}
\Omega(t)= & -\frac{1}{2}\frac{d}{dt}\ln\left(1+\left(\frac{X_{1}\left(t\right)}{X_{2}\left(t\right)}\right)^{2}\right),\\
\Gamma_{d}(t)= & \frac{1}{4}\frac{d}{dt}\ln\left(\frac{Z_{1}\left(t\right)}{X_{1}^{2}\left(t\right)+X_{2}^{2}\left(t\right)}\right),\\
\Gamma_{-}(t)= & -\frac{1}{2}\left[\frac{dZ_{2}\left(t\right)}{dt}+\frac{d\ln Z_{1}\left(t\right)}{dt}\left(1-Z_{2}\left(t\right)\right)\right],\\
\Gamma_{+}(t)= & -\frac{1}{2}\left[-\frac{dZ_{2}\left(t\right)}{dt}+\frac{d\ln Z_{1}\left(t\right)}{dt}\left(1+Z_{2}\left(t\right)\right)\right].
\end{align}
The first term at the right-hand side of Eq.\,\eqref{ME} corresponds
to the unitary evolution, and the other three terms represent the
dephasing, dissipation and absorption processes with rates $\Gamma_{d}(t),\,\Gamma_{-}(t)$,
and $\Gamma_{+}(t)$, respectively. The negativity of the rates can
serve as a proper indicator of non-Markovianity of the system dynamics,
closely relating to the functions $X_{1}\left(t\right)$, $X_{2}\left(t\right)$,
$Z_{1}\left(t\right)$ and $Z_{2}\left(t\right)$.

In the following sections, we will investigate the non-Markovianity
of the central spin dynamics in a more intuitive way, in terms of
the information flow between the system and the bath, quantified by
the change of the trace distance between two states of the central
spin which also depends on the above functions.

\section{NON-MARKOVIANITY of system dynamics\label{sec:DYNAMICS-OF-NON-MARKOVIANITY}}

Now we use the information backflow to quantify the non-Markovianity
of the quantum dynamics of the central spin interacting with bath
spins. We consider two different pairs of initial states $\left\{ \lvert0\rangle,\lvert1\rangle\right\} $
and $\left\{ \lvert+\rangle,\lvert-\rangle\right\} $ to compute the
trace distance. It is shown below that the trace distance between
two arbitrary initial states can be decomposed into the trace distances
of these two initial state pairs.

To facilitate the computation of the trace distance, we assume that
the number of bath qubits $N$ is sufficiently large but finite. We
will show the effect of the number of bath spins, i.e.\textcolor{black}{{}
the dimension of the bath,} \textcolor{black}{as well as} the coupling
strength, the frequency detuning and the bath temperature,\textcolor{black}{{}
on the information backflow.} The results turn out to show collapse-revival
patterns in the information backflow, and that the periods and amplitudes
of the collapse-revival patterns may characterize the non-Markovianity
of the system dynamics while the usual integration of the trace distance
increase over the evolution time may diverge.

\subsection{Trace distance given two initial states}

The trace distance between two quantum states $\rho_{1}$ and $\rho_{2}$
is defined in Eq.\,\eqref{td}, which can be recast into a much more
intuitive form in terms of Bloch vectors \citep{nielsen20102},
\begin{equation}
D\left(t\right)=\frac{1}{2}\left|\boldsymbol{v}_{1}-\boldsymbol{v}_{2}\right|,\label{tdof2}
\end{equation}
where $\boldsymbol{v}_{1},\boldsymbol{v}_{2}$ are the Bloch vectors
of $\rho_{1},\rho_{2}$ respectively and $|\cdot|$ denotes the Euclidean
distance.

In the current problem, the Bloch vectors $\boldsymbol{v}_{1},\boldsymbol{v}_{2}$
of the central spin are time-dependent as derived in Eq.\,\eqref{reducedsys},
and the trace distance between two states of the central spin at time
$t$ given arbitrary initial states $\rho_{1}(0),\,\rho_{2}(0)$ can
be obtained as
\begin{equation}
D\left(t\right)=\frac{1}{2}\sqrt{\alpha_{z}\tds[z]+\alpha_{x}\tds},\label{tds}
\end{equation}
where the coefficients $\alpha_{x}=[x_{1}(0)-x_{2}(0)]^{2}+[y_{1}(0)-y_{2}(0)]^{2}$
and $\alpha_{z}=[z_{1}(0)-z_{2}(0)]^{2}$ are determined by the two
initial states, and the functions $\td$ and $\td[x]$ represent the
trace distance given the initial states $|0\rangle,|1\rangle$ and
given the initial states $|\pm\rangle$, respectively. It is straightforward
to verify that
\begin{align}
\td= & \left|Z_{1}\left(t\right)\right|,\label{D01}\\
\td[x]= & \sqrt{X_{1}^{2}\left(t\right)+X_{2}^{2}\left(t\right)}.\label{Dpm}
\end{align}
It can be seen that the trace distance given two arbitrary initial
states can be determined by $\td$ and $\td[x]$; therefore, we will
only study the time evolution of $\td$ and $\td[x]$ in the following.

\subsection{Dynamics of information backflow}

Now we study the time evolution of the information flow. To derive
analytical results for the trace distance of the central spin, we
need to carry out the summations in Eq.\,\eqref{eq:x1t}-\eqref{eq:z2t},
which are quite complex. To simplify the computation, we assume the
number of bath spins $N$ to be much larger than $T/\omega_{b}$ but
still finite so that the terms $e^{-\frac{M\omega_{b}}{T}}$ associated
with the Dicke states $|J,M\rangle$ in the thermal state of the bath
spins will have negligible contribution when $M$ is large. For the
convenience of the computation, we replace $M$ in the Dicke state
$|J,M\rangle$ with a renormalized parameter $\xi=\left(M+J\right)/N$,
so that $\xi$ will range from 0 to 1 with a fixed small step $1/N$,
and the trace distance can be expanded to the first few lower orders
of $\xi$.

\subsubsection{Trace distance $\protect\td$\label{subsec:Trace-distance z}}

For the trace distance $\td$, it only depends on the parameter $Z_{1}\left(t\right)$
in Eq.\,\eqref{eq:z1t}. We can expand $Z_{1}\left(t\right)$ to
a Taylor series,

\begin{equation}
\begin{aligned}Z_{1}\left(t\right)= & 1-\left(1+e^{\frac{\omega_{b}}{T}}\right)\sum_{\xi}\frac{e^{-\frac{\xi J\omega_{b}}{T}}}{Q}\\
 & \times\sum_{j}a_{j}\xi^{j}\left[1-\cos\left(\sum_{k}\nu_{k}\xi^{k}t\right)\right],
\end{aligned}
\label{Z1t}
\end{equation}
and work out the summation up to $O(1/N^{2})$. Applying it to Eq.\,\eqref{D01},
the trace distance $\td$ can be approximately obtained as (see Appendix~\ref{sec:the-approximation-of})

\begin{equation}
\td\doteq\overline{D_{z}}-W_{z}\left(t\right)\cos\left(\nu_{0}t-\phi_{3}\right),\label{D01a}
\end{equation}
where the mean $\overline{D_{z}}$ is independent of the time $t$
and given as
\begin{equation}
\overline{D_{z}}=1-\frac{a_{1}}{N}\coth\frac{\omega_{b}}{2T}-\frac{a_{2}}{N^{2}}\coth^{2}\frac{\omega_{b}}{2T},
\end{equation}
and the amplitude of the oscillation around the mean is
\begin{align}
W_{z}\left(t\right)= & \frac{\sinh\frac{\omega_{b}}{T}}{P_{-}\left(\frac{4g\Delta}{\nu_{0}}t\right)}\left[\frac{a_{1}^{2}}{N^{2}}+\frac{a_{2}^{2}}{N^{4}}\frac{P_{+}\left(\frac{4g\Delta}{\nu_{0}}t\right)}{P_{-}\left(\frac{4g\Delta}{\nu_{0}}t\right)}\right.\nonumber \\
 & \left.+2\frac{a_{1}a_{2}}{N^{3}}\sqrt{\frac{P_{+}\left(\frac{4g\Delta}{\nu_{0}}t\right)}{P_{-}\left(\frac{4g\Delta}{\nu_{0}}t\right)}}\cos\left(\phi_{1}-\phi_{2}\right)\right]^{1/2},
\end{align}
where the phases $\phi_{1},\phi_{2}$ are defined in Appendix \ref{sec:the-approximation-of},
and the function $P\left(x\right)$ is defined as
\begin{equation}
P_{\pm}\left(x\right)=\cosh\frac{\omega_{b}}{T}\pm\cos x.
\end{equation}
The coefficients $a_{1},a_{2}$ are

\begin{equation}
\begin{aligned}a_{1} & =\frac{8g^{2}N(N+1)}{\nu_{0}^{2}},\\
a_{2} & =-\frac{8g^{2}N^{2}(\nu_{0}+2\Delta)^{2}}{\nu_{0}^{4}},
\end{aligned}
\label{A1A2}
\end{equation}
and $\nu_{0}$ is
\begin{equation}
\nu_{0}=2(N+1)g-\Delta.\label{eq:v0}
\end{equation}

Before continuing the computation, let us pause and have a digestion
of the result in Eq.\,\eqref{D01a}. We can see that $\td$ is a
combination of two oscillations: one is a rapid oscillation in the
cosine term $\cos\left(\nu_{0}t+\phi_{3}\right)$ with the frequency
$\nu_{0}$ given in Eq.\,\eqref{eq:v0} which is of  $O(N)$, and
the other is a slow oscillation with the frequency $\frac{4g\Delta}{\nu_{0}}$
which is of $O(N^{-1})$ in the terms $P_{\pm}\left(\frac{4g\Delta}{\nu_{0}}t\right)$.
This implies that the amplitude of the rapid oscillation will change
slowly but periodically with time, and a ``collapse-revival'' phenomenon
will appear in $\td$, which is similar to the collapse-revival phenomenon
in quantum optics, i.e., the collapse-revival of the atomic population
inversion when a two-level atom is interacting with a single mode
bosonic field. We denote the frequency of the collapse-revival patterns
as
\begin{equation}
\nu_{cr}=\frac{4g\Delta}{\nu_{0}},\label{eq:vcr}
\end{equation}
and will find it is universal for the collapse-revival phenomenon
with arbitrary initial states of the bath.

\begin{figure}
\includegraphics[width=8cm]{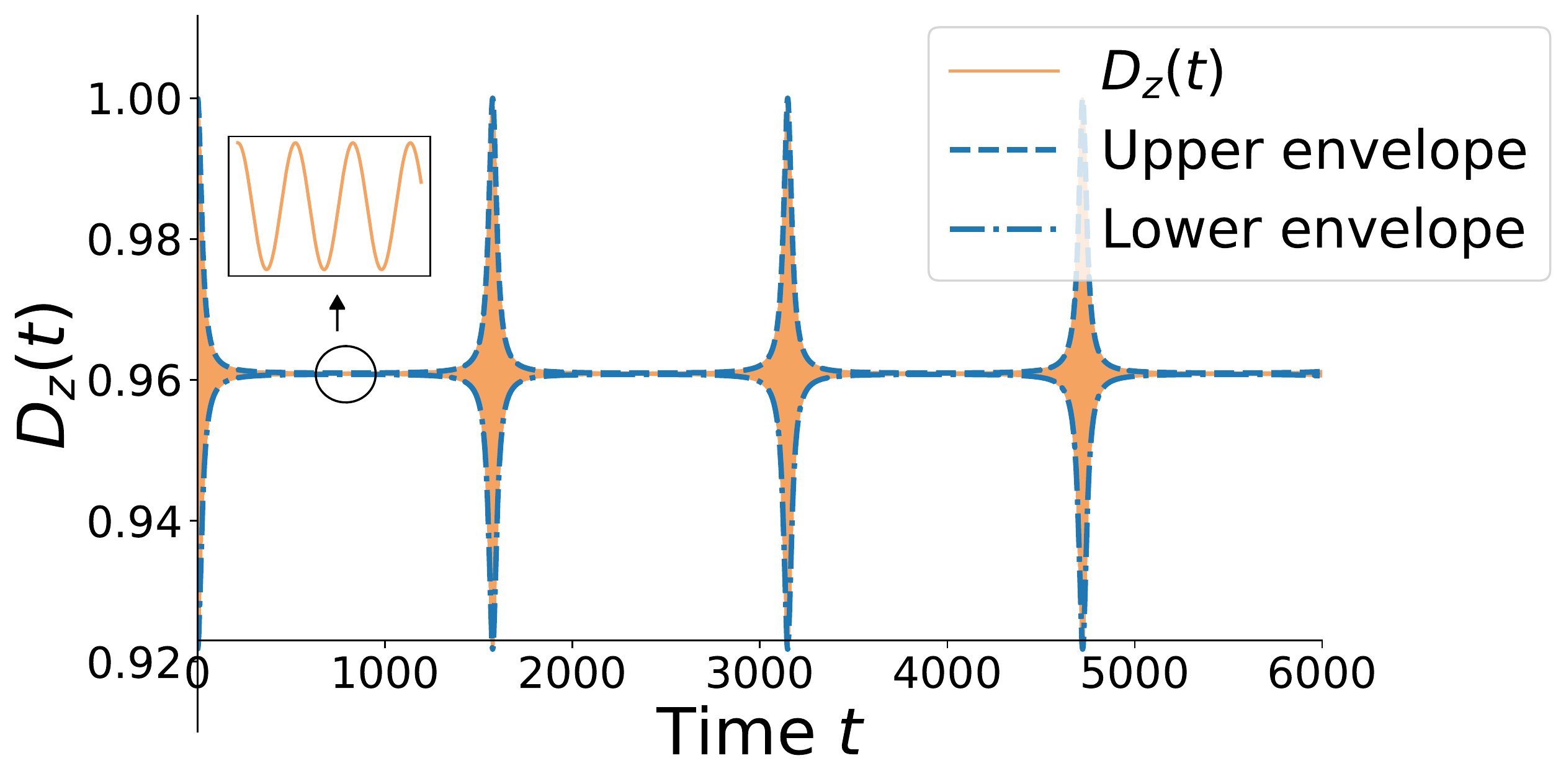}

\caption{Time evolution and envelope of $\protect\td$ in a long time scale.
The envelope consists of an upper line and a lower line. The time
evolution is plotted by exact numerical computation, and the envelope
lines are plotted according to the analytical result in Eq.\,\eqref{D01enve}
with signs $+$ and $-$ respectively. The rapid oscillation with
a frequency $\nu_{0}$ is shown as the sinusoidal solid line in the
zoomed panel. It can be observed that the amplitude of $\protect\td$
is periodic in a long time scale, showing a collapse-revival pattern
and implying non-vanishing information flow between the central spin
and the bath spins. Parameters: $N=1000$, $g=1$, $\omega_{s}=3$,
$\omega_{b}=1$, and $T=10$.}
\label{fig:D01s}
\end{figure}

To have an intuitive picture of this phenomenon, the trace distance
$\td$ and its envelopes are plotted in Fig.\,\ref{fig:D01s}. In
the figure, one can observe that the amplitude of $\td$ decreases
rapidly to almost zero first and stays for a while, then the oscillation
revives and the amplitude of $\td$ increase to almost the original
value again, and such a process will repeat. This phenomenon is essentially
rooted in the superposition of oscillations with different frequencies
where the phases of different oscillations will match and mismatch
periodically with time.

To give a detailed analysis of the collapse-revival pattern in $\td$,
we derive the envelopes of $\td$ by taking the amplitude of the rapid
oscillation, and the result turns out to be

\begin{equation}
\Gamma_{z}(t)=\overline{D_{z}}\pm W_{z}\left(t\right),\label{D01enve}
\end{equation}
where the $+$ and $-$ signs represent the upper and lower envelopes
respectively.

The oscillation term $W_{z}\left(t\right)$ has a maximum value
\begin{equation}
\max W_{z}\left(t\right)=\frac{a_{1}}{N}\coth\frac{\omega_{b}}{2T}+\frac{a_{2}}{N^{2}}\coth^{2}\frac{\omega_{b}}{2T},\label{eq:wzt}
\end{equation}
and thus
\begin{equation}
\overline{D_{z}}+\max W_{z}\left(t\right)=1,
\end{equation}
implying that the trace distance $\td$ can almost return to its initial
value, which indicates that most information can flow back to the
system from the bath and there is no irreversible dissipation of the
information in this scenario.

Note the effect of the number of bath spins $N$ in the collapse-revival
pattern: when $N$ becomes larger, the amplitude of the collapse-revival
pattern $W_{z}\left(t\right)$ will be smaller, which means the information
backflow between the system and the bath will decrease, indicating
a weaker non-Markovianity. This shows how the non-Markovian dynamics
transits to Markovian dynamics with an increasing dimension of the
bath from one aspect. As we will see below, the period of the collapse-revival
pattern can also show the effect of an increasing $N$ on the transition
of the Markovianity, from another aspect. We also note the different
roles of the frequency detuning $\Delta=\omega_{s}-\omega_{b}$ and
the bath frequency $\omega_{b}$ as well as the bath temperature $T$
in the collapse-revival pattern of the trace distance $\td$: $\Delta$
determines the period of collapse-revival pattern,
\begin{equation}
T_{cr}=\frac{2\pi}{\nu_{cr}}=\frac{\pi[2(N+1)g-\Delta]}{2g\Delta},\label{eq:tcr}
\end{equation}
while the bath frequency $\omega_{b}$ and the temperature $T$ affect
the amplitude of the collapse-revival pattern via the exponents $e^{\frac{\omega_{b}}{T}}$,
$e^{\frac{2\omega_{b}}{T}}$, etc.

In addition, note that in this case, the average of the trace distance
does not change with time, so the envelope is mainly determined by
the oscillation amplitudes of the trace distance. This will be in
sharp contrast to the behavior of the trace distance $D_{x}(t)$ in
the following.

\subsubsection{Trace distance $\protect\td[x]$\label{subsec:Trace-distance x}}

For the trace distance $\td[x]$, we also keep the terms up to $O(1/N^{2})$.
The trace distance turns out to be
\begin{equation}
\tds[x]\doteq\overline{D_{x}^{2}}\left(t\right)+W_{x}^{2}\left(t\right)\cos\left(\nu_{0}t-\phi_{4}\right),\label{eq:Dpma}
\end{equation}
where the mean value and the amplitude of collapse-revival pattern
are given as (see Appendix \ref{sec:the-approximation-of-1} for the
derivation)
\begin{align}
\overline{D_{x}^{2}}\left(t\right)= & \frac{\left(\cosh\frac{\omega_{b}}{T}-1\right)\left(1-\frac{a_{3}}{N}\right)^{2}}{P_{-}\left(\frac{4g\Delta}{\nu_{0}}t\right)}+\frac{a_{3}^{2}P_{+}\left(\frac{4g\Delta}{\nu_{0}}t\right)}{N^{2}\left(\cosh\frac{\omega_{b}}{T}-1\right)}\nonumber \\
 & +\frac{2\text{\ensuremath{a_{3}}}\left(\cosh\frac{\omega_{b}}{T}-1\right)\left[e^{-\frac{\omega_{b}}{T}}-\left(1-\frac{a_{3}}{N}\right)\cos\left(\frac{4g\Delta}{\nu_{0}}t\right)\right]}{NP_{-}^{2}\left(\frac{4g\Delta}{\nu_{0}}t\right)},\label{AX2}\\
W_{x}^{2}\left(t\right)= & 2\frac{a_{3}}{N}\left[\frac{P_{+}\left(\frac{4g\Delta}{\nu_{0}}t\right)}{P_{-}\left(\frac{4g\Delta}{\nu_{0}}t\right)}\right]^{1/2},\label{eq:Wx}
\end{align}
where the coefficient $a_{3}$ is
\begin{equation}
a_{3}=\frac{4g^{2}N^{2}}{\left(2gN-\Delta\right)^{2}}.
\end{equation}
The envelope of $D_{x}(t)$ can be obtained directly,
\begin{equation}
\Gamma_{x}\left(t\right)=\sqrt{\overline{D_{x}^{2}}\left(t\right)\pm W_{x}^{2}\left(t\right)}.\label{eq:envelop x}
\end{equation}
Note that the mean value of $\tds[x]$ is dependent on the time $t$
in this case, in contrast to the case of $D_{z}(t)$.

Similar to $\td$, the trace distance $\td[x]$ includes two oscillations,
a rapid oscillation with frequency $\text{\ensuremath{\nu_{0}}}$
of $O(N)$ and a slow oscillation with frequency $\frac{4g\Delta}{\nu_{0}}$
of $O(N^{-1})$. Due to the combination of two oscillations with different
frequencies, the amplitude of the rapid oscillation has a periodic
collapse-revival pattern repeated with the smaller frequency $\frac{4g\Delta}{\nu_{0}}$.
But in contrast to the case of $D_{z}(t)$, the average of $D_{x}(t)$
changes significantly with time while the amplitude of the rapid oscillation
is only of order $N^{-1}$, so the envelope of the trace distance
is mainly determined by the average in this case.

\begin{figure}[h]
\subfloat[\centering]{\includegraphics[width=8cm]{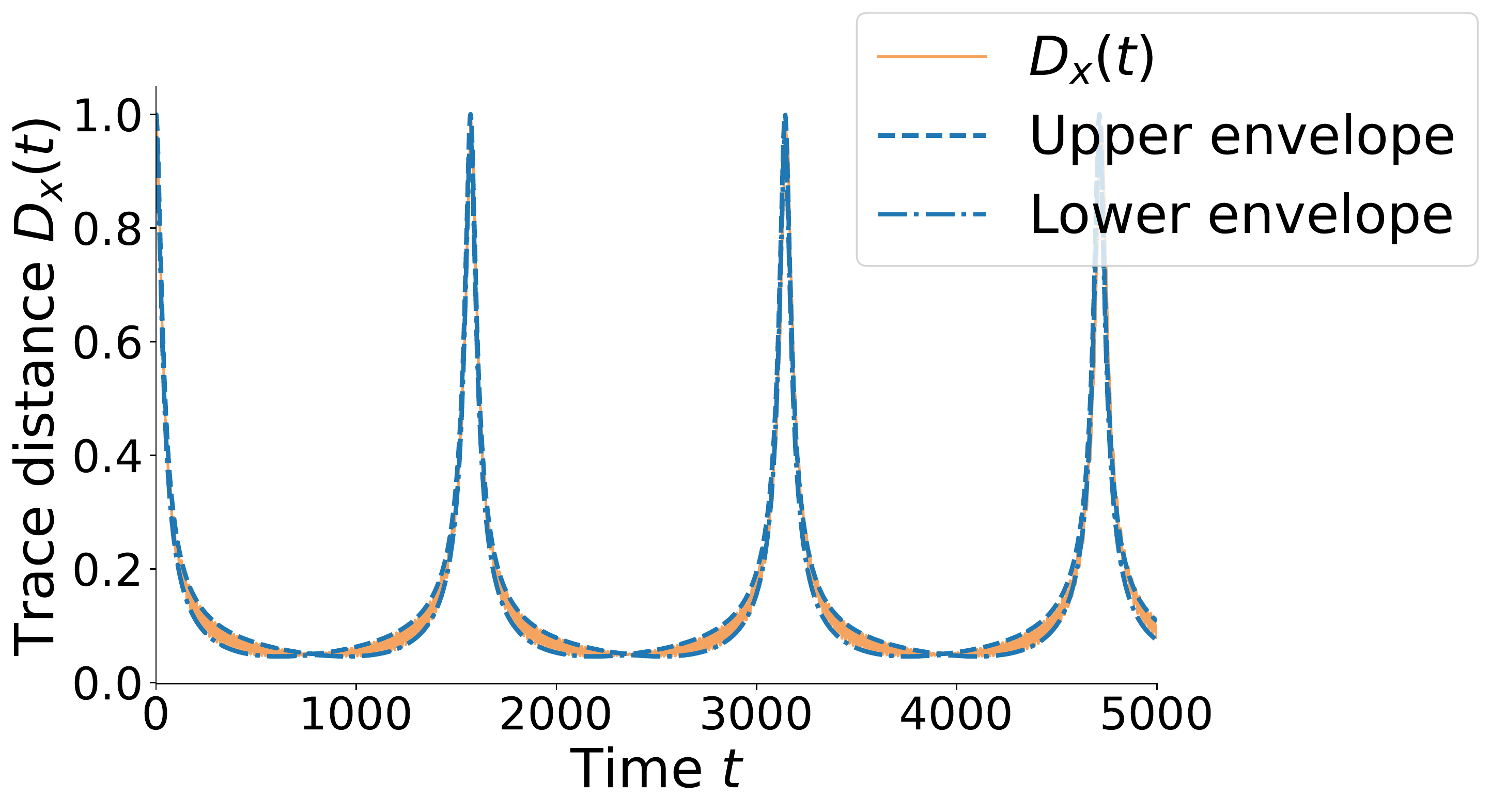}

}

\subfloat[\centering]{\includegraphics[width=8cm]{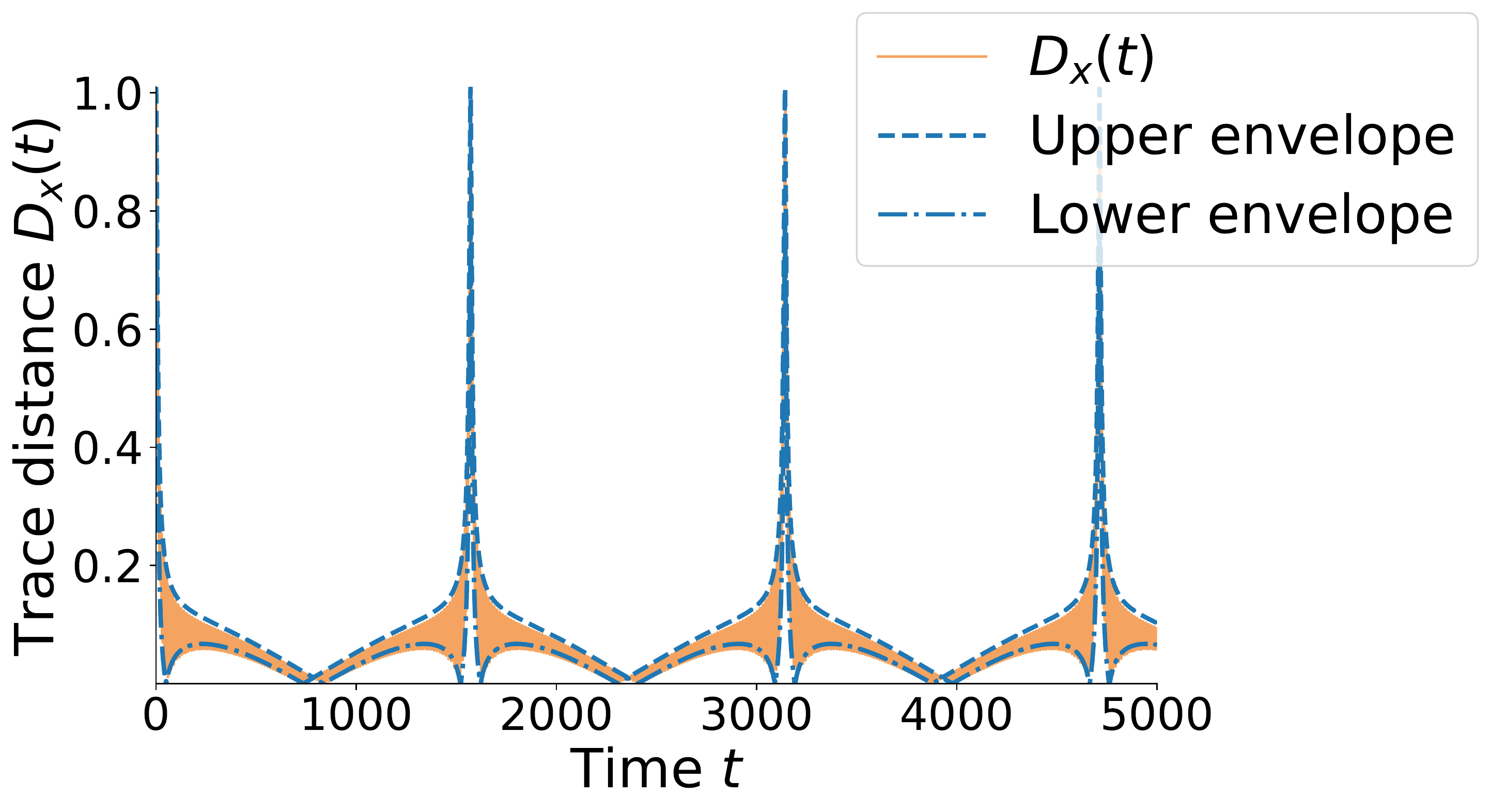}

}\caption{Time evolution and envelope of $\protect\td[x]$ in a long time scale
with different bath temperatures: (a) $T=10$ and (b) $T=50$, showing
a periodic collapse-revival pattern with two envelope lines. The time
evolution is plotted by exact numerical computation, and the envelope
lines are plotted according to the analytical result in Eq.\,\eqref{eq:envelop x}
with signs $+$ and $-$ respectively. Parameters: $\omega_{s}=3,$
$N=1000$, $g=1$, and $\omega_{b}=1$.}
\label{fig:Dpmbs}
\end{figure}

Figure \ref{fig:Dpmbs} shows the time evolution and envelope of $\td[x]$
with different $\frac{\omega_{b}}{T}$. It can be observed that the
envelope consists of two lines above and below the average, and the
amplitude of the trace distance exhibits a collapse-revival pattern
with frequency $\frac{4g\Delta}{\nu_{0}}$, the same as $\nu_{cr}$
in $\td[z]$. The two envelope lines are very close to the average
of the trace distance, which verifies the results above.

It can be verified straightforwardly that when the number of bath
spins $N$ goes to infinity asymptotically, the average of the squared
trace distance $\overline{D_{x}^{2}}\left(t\right)$ will approach
$1$ and the fluctuation $W_{x}^{2}\left(t\right)$ will vanish, implying
the information of the system keeps almost unchanged over the evolution
time and the information flow between the system and the bath is significantly
suppressed in this limit, which indicates a weaker non-Markovianity
of the system dynamics.

\subsubsection{Trace distances between arbitrary states of central spin\label{subsec:Trace-distances-between}}

From the above results, one can see that the frequencies of the collapse-revival
patterns for the trace distances $D_{x}(t)$ and $D_{z}(t)$ are the
same, and Eq.\,\eqref{tds} shows that the trace distance between
two arbitrary states of the central spin can be determined by $D_{x}(t)$
and $D_{z}(t)$, so it can be immediately concluded that there exists
collapse-revival structure in the trace distance between arbitrary
bath spin states if it exists for the pairs of bath states $|0\rangle,|1\rangle$
or $|\pm\rangle$, given the number of the bath spins, the system-bath
interaction strength and frequency detuning. It can be obtained from
Eq.\,\eqref{tds} that in general when the number of bath spins,
$N$, is large, the trace distance between the central spin states
evolved from two arbitrary initial states is
\begin{equation}
D^{2}\left(t\right)=\overline{D^{2}}\left(t\right)+W^{2}\left(t\right)\cos\left(\nu_{0}t-\phi\right),
\end{equation}
where its mean value and the amplitude of oscillation are
\begin{align}
\overline{D^{2}\left(t\right)}= & \frac{1}{4}\alpha_{z}\overline{D_{z}}^{2}+\frac{1}{4}\alpha_{x}\overline{D_{x}^{2}}\left(t\right),\\
W^{2}\left(t\right)= & \frac{1}{2NP_{-}\left(\nu_{cr}t\right)}\left[\alpha_{z}^{2}a_{1}^{2}\sinh^{2}\frac{\omega_{b}}{T}+\alpha_{x}^{2}a_{3}^{2}P\left(\nu_{cr}t\right)\right.\nonumber \\
 & \left.-2\alpha_{x}\alpha_{z}a_{1}a_{3}\sinh\frac{\omega_{b}}{T}P^{\frac{1}{2}}\left(\nu_{cr}t\right)\cos\left(\phi_{3}-\phi_{4}\right)\right]^{1/2},
\end{align}
and the phase of the oscillation is 
\begin{equation}
\phi=\arctan\frac{\alpha_{z}a_{1}\sinh\frac{\omega_{b}}{T}\sin\phi_{3}-\alpha_{x}a_{3}P^{\frac{1}{2}}\left(\nu_{cr}t\right)\sin\phi_{4}}{\alpha_{z}a_{1}\sinh\frac{\omega_{b}}{T}\cos\phi_{3}-\alpha_{x}a_{3}P^{\frac{1}{2}}\left(\nu_{cr}t\right)\cos\phi_{4}},
\end{equation}
where $P(x)$ is defined as
\begin{equation}
P(x)=P_{+}(x)P_{-}(x)=\cosh^{2}\frac{\omega_{b}}{T}-\cos^{2}x.
\end{equation}
The envelope of $D(t)$ can be obtained directly,
\begin{equation}
\Gamma_{\pm}\left(t\right)=\sqrt{\overline{D^{2}}\left(t\right)\pm W^{2}\left(t\right)}.\label{eq:evn}
\end{equation}
The mathematical detail of the derivation is left in Appendix \ref{sec:Trace-distance-for}.

This suggests that the collapse-revival structure exists universally
for different pairs of the central spin states, sharing the same collapse-revival
frequency $\nu_{cr}$ in Eq.\,\eqref{eq:vcr} but differing drastically
in the behavior of the mean and the amplitude of the fast oscillation
of the information flow.

\begin{figure}
\subfloat[\centering]{\includegraphics[width=8cm]{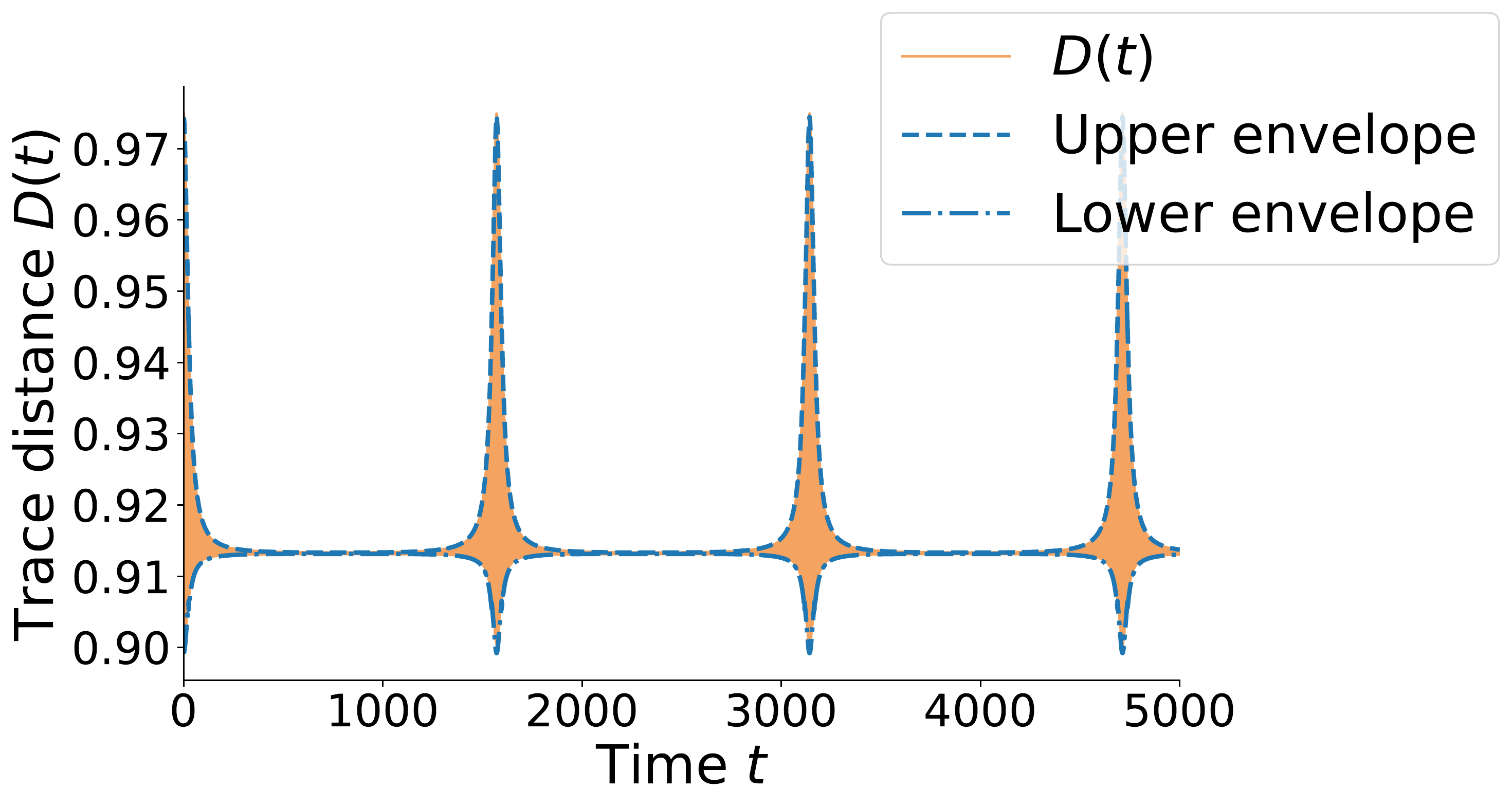}

}

\subfloat[\centering]{\includegraphics[width=8cm]{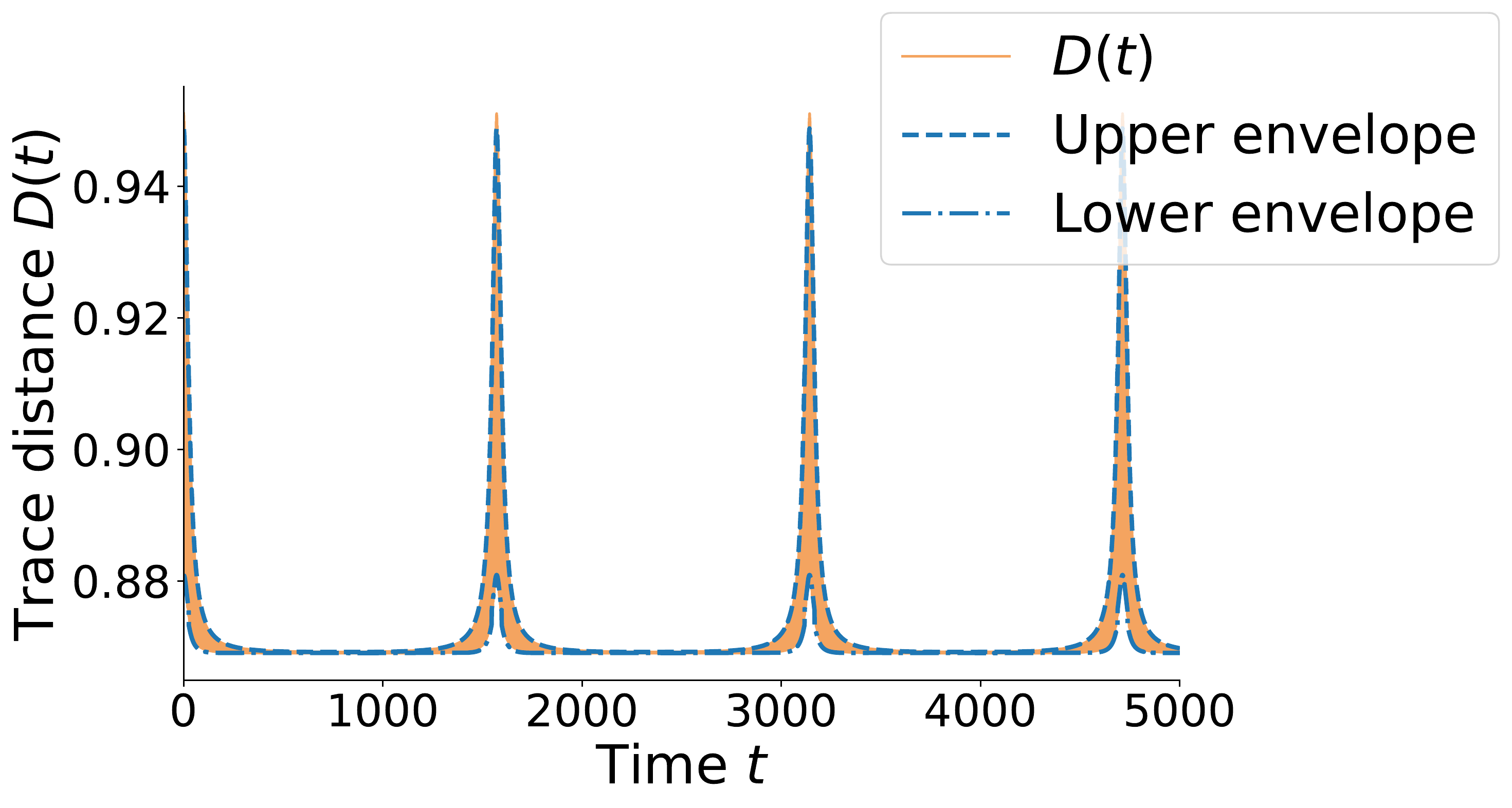}}

\caption{Time evolution and envelope of $D\left(t\right)$ with coefficients
(a) $\alpha_{z}=3.614$ and $\alpha_{x}=0.188$, (b) $\alpha_{z}=3.273$
and $\alpha_{x}=0.345$. While the upper envelope always has upward
peaks, the lower envelope can have upward or downward peaks: as $\alpha_{x}$
increases and $\alpha_{z}$ decreases, a downward peak of the lower
envelope will move up and finally become an upward peak. Parameters:
$N=1000$, $g=1$, $\omega_{s}=3$, $\omega_{b}=1$, and $T=10$.}
\label{fig:Dtarb}
\end{figure}

Fig.\,\ref{fig:Dtarb} illustrates the above results for a pair of
non-orthogonal system states numerically.

\begin{figure}[h]
\subfloat{\includegraphics[width=7cm]{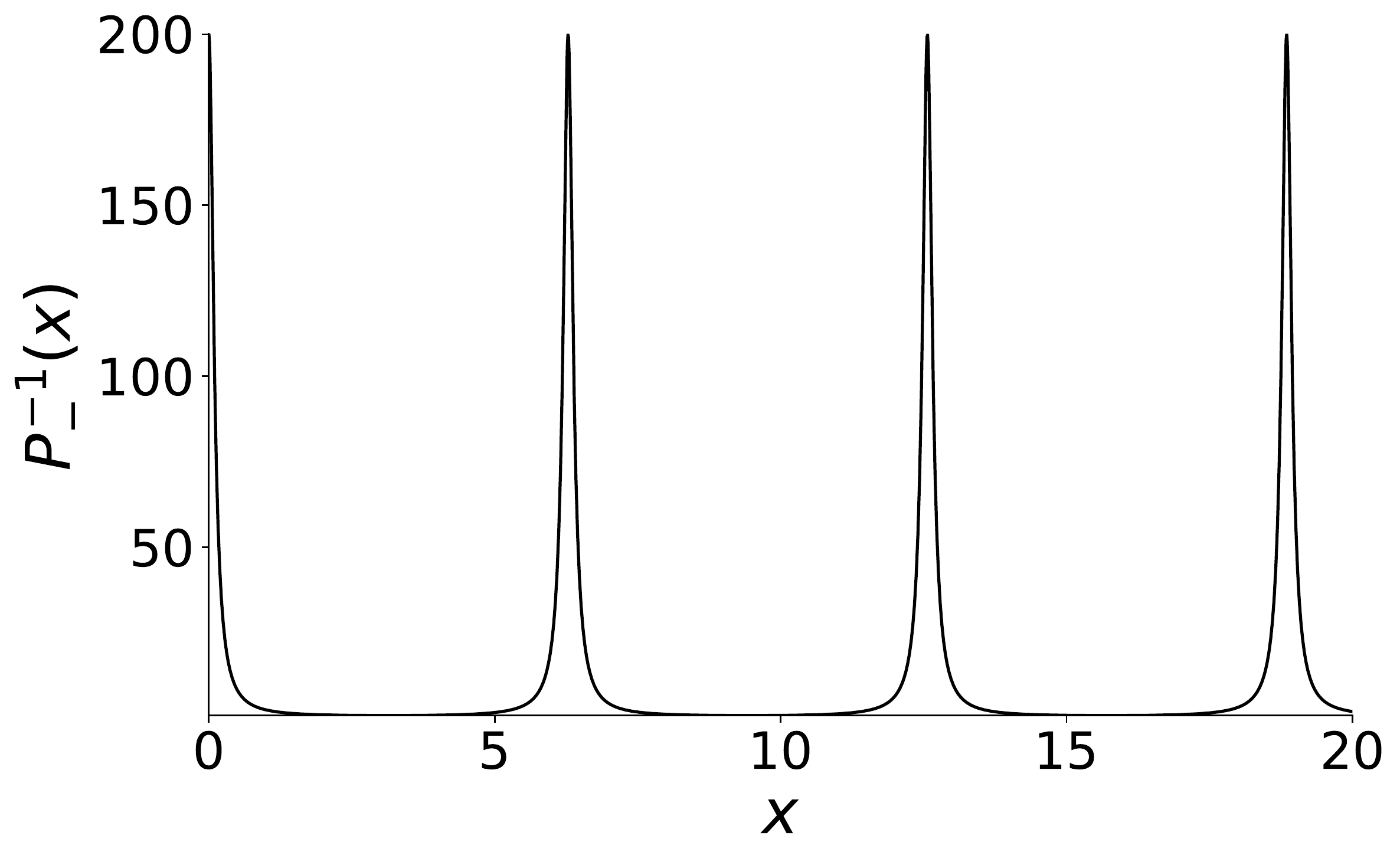}}

\caption{The behavior of the function $P_{-}^{-1}(x)$. It has large flat regions
separated with periodic peaks, which determines the collapse-revival
pattern. Parameters: $\omega_{b}=1$, and $T=10$.}
\label{fig:Px}
\end{figure}

To capture the main feature of the collapse-revival pattern, we note
that for $D_{z}(t)$ the mean does not change over time and approaches
$1$ and the oscillation is mainly determined by the function $P_{-}^{-1}\left(\nu_{cr}t\right)$
if only the lowest order term is considered, while for $D_{x}(t)$
the function $P_{-}^{-1}\left(\nu_{cr}t\right)$ dominates the mean
value ranging from $0$ to $1$, and the amplitude of the oscillation
with $O(N^{-1})$ is small compared to the mean value. So, the behavior
of the general trace distance $D\left(t\right)$ can be simplified
\begin{equation}
\Gamma_{\pm}(t)\doteq\sqrt{\frac{\alpha_{z}}{4}\overline{D_{z}}^{2}+\frac{\alpha_{x}}{4}\frac{\cosh\frac{\omega_{b}}{T}-1}{P_{-}\left(\nu_{cr}t\right)}\pm\frac{\alpha_{z}a_{1}}{2N}\frac{\sinh\frac{\omega_{b}}{T}}{P_{-}\left(\nu_{cr}t\right)}},\label{eq:enve}
\end{equation}
where only the lowest order terms of $1/N$ in $D_{x}(t)$ and $D_{z}(t)$
remain, respectively, and the function $P_{-}^{-1}\left(\nu_{cr}t\right)$
determines the time evolution of the trace distance. In fact, it is
exactly the function $P_{-}^{-1}(x)$ that the collapse-revival pattern
stems from, as $P_{-}^{-1}(x)$ has large flat regions separated by
periodic sharp peaks, plotted in Fig.\,\ref{fig:Px}.

An interesting observation of $\Gamma_{\pm}(t)$ is that while there
is only one type of upper envelope $\Gamma_{+}(t)$ which has an upward
peak, there are two types of lower envelopes $\Gamma_{-}(t)$ with
an upward peak and a downward peak respectively. This depends on the
competition between the $\alpha_{x}$ term and the $\alpha_{z}$ term
in Eq.\,\eqref{eq:enve}: if the $\alpha_{x}$ term is larger than
the $\alpha_{z}$ term, the coefficient of $P_{-}^{-1}\left(\nu_{cr}t\right)$
is positive and the peak of $\Gamma_{-}(t)$ is upward, otherwise
the coefficient of $P_{-}^{-1}\left(\nu_{cr}t\right)$ becomes negative
and the peak of $\Gamma_{-}(t)$ turns to be downward accordingly.
This is in accordance with the two types of envelopes shown in Fig.
\eqref{fig:Dtarb}.

\section{Characteristic time scales of collapse-revival patterns\label{sec:Characteristic-time-scales}}

In the previous section, we obtained the behavior of information flow
between the central spin and the bath, and showed the existence of
the collapse-revival structure in the trace distance for arbitrary
states of the central spin. To describe the collapse-revival phenomenon
more quantitatively, we study typical characteristic time scales of
the collapse-revival patterns in detail in this section. The time
scales we consider include the period of the collapse-revival pattern,
the collapse time and the revival time. We will obtain analytical
results for these time scales and analyze the roles of the interaction
strength, the frequency detuning, etc. in these times scales. In particular,
we will consider how the number of bath qubits affects these time
scales, in order to show the role of the bath dimension on the Markovianity
of quantum dynamics.

\subsection{Various time scales of collapse-revival pattern}

The periodicity is the most prominent characteristic of the collapse-revival
pattern, so we study the period of the collapse-revival pattern first.

It has been shown above that the frequency of the collapse-revival
pattern is always $\nu_{cr}=\frac{4g\Delta}{\nu_{0}}$ for central
spin states evolved from arbitrary initial states, so the period of
the collapse-revival pattern is
\begin{equation}
T_{cr}=\frac{2\pi}{\nu_{cr}}=\frac{\pi(N+1)}{\Delta}-\frac{\pi}{2g}.\label{eq:tc}
\end{equation}
When $g$ is not small or $N$ is sufficiently large so that $gN\gg\Delta$,
$T_{cr}$ can be simplified to
\begin{equation}
T_{cr}\doteq\frac{\pi N}{\Delta},
\end{equation}
implying that the period increases with a larger $N$ or a smaller
$\Delta$.

\begin{figure}[h]
\subfloat[\centering]{\includegraphics[width=4.2cm]{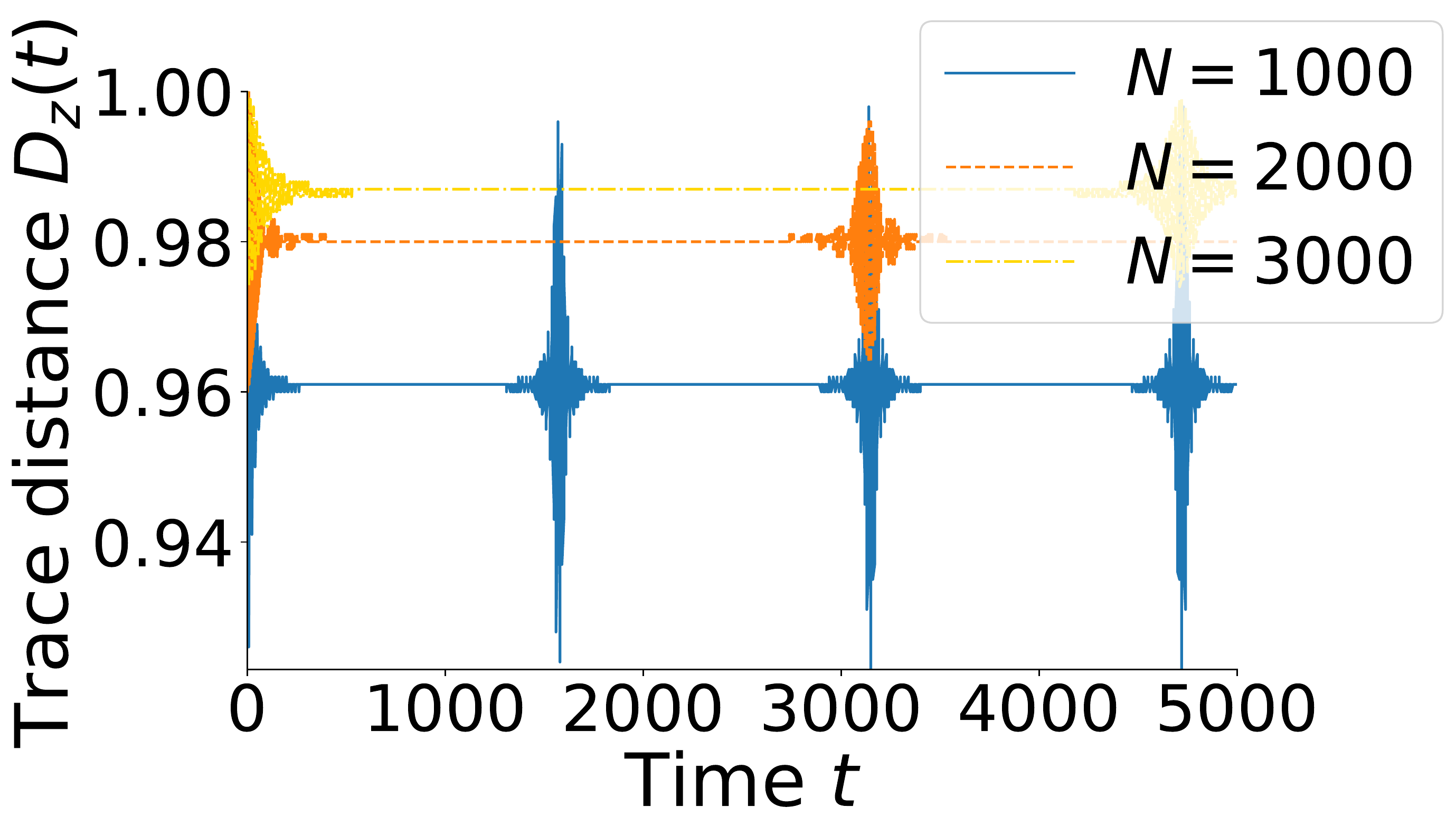} }\subfloat[\centering]{\includegraphics[clip,width=4.2cm]{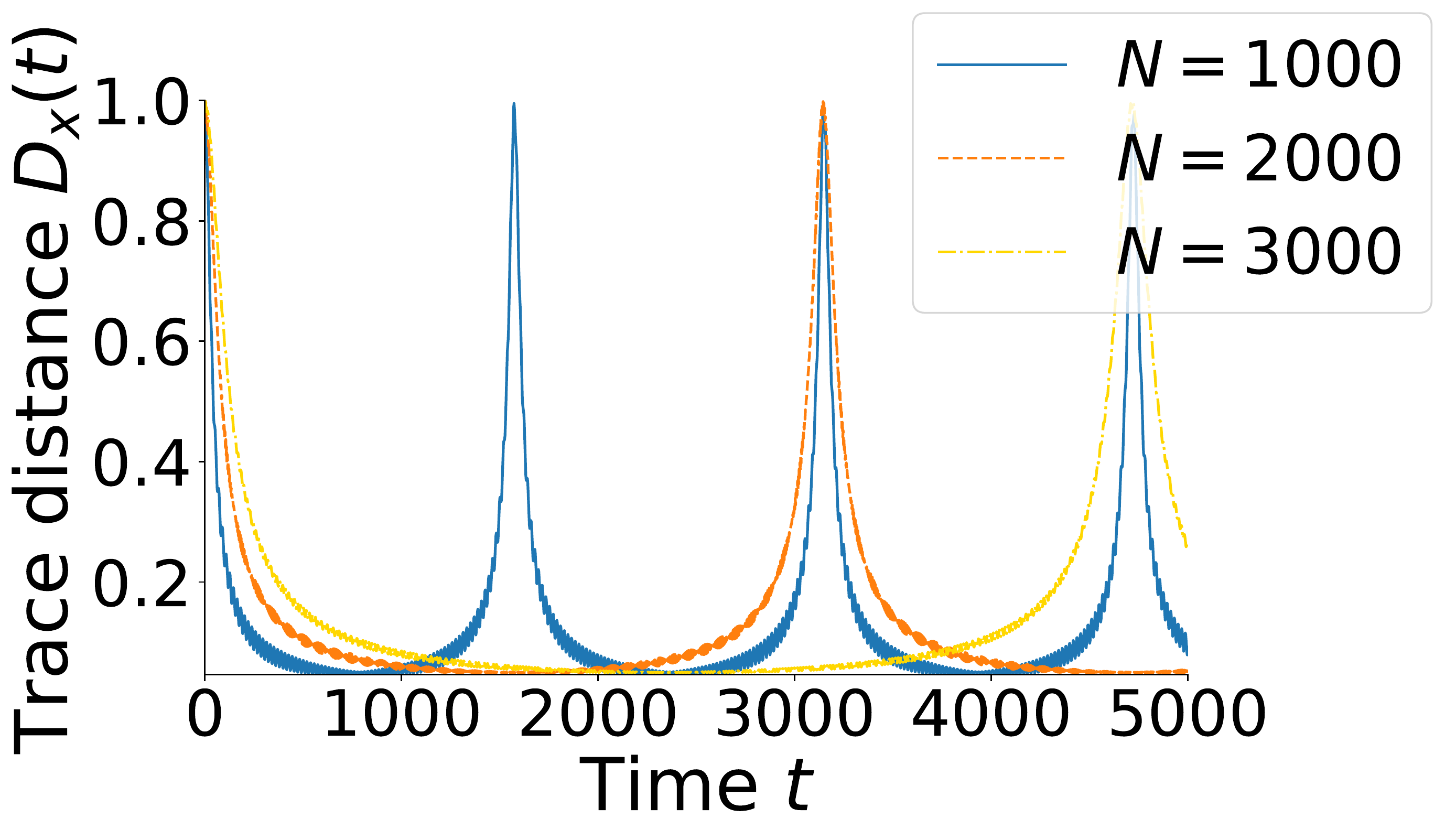}

}

\subfloat[\centering]{\includegraphics[width=4.2cm]{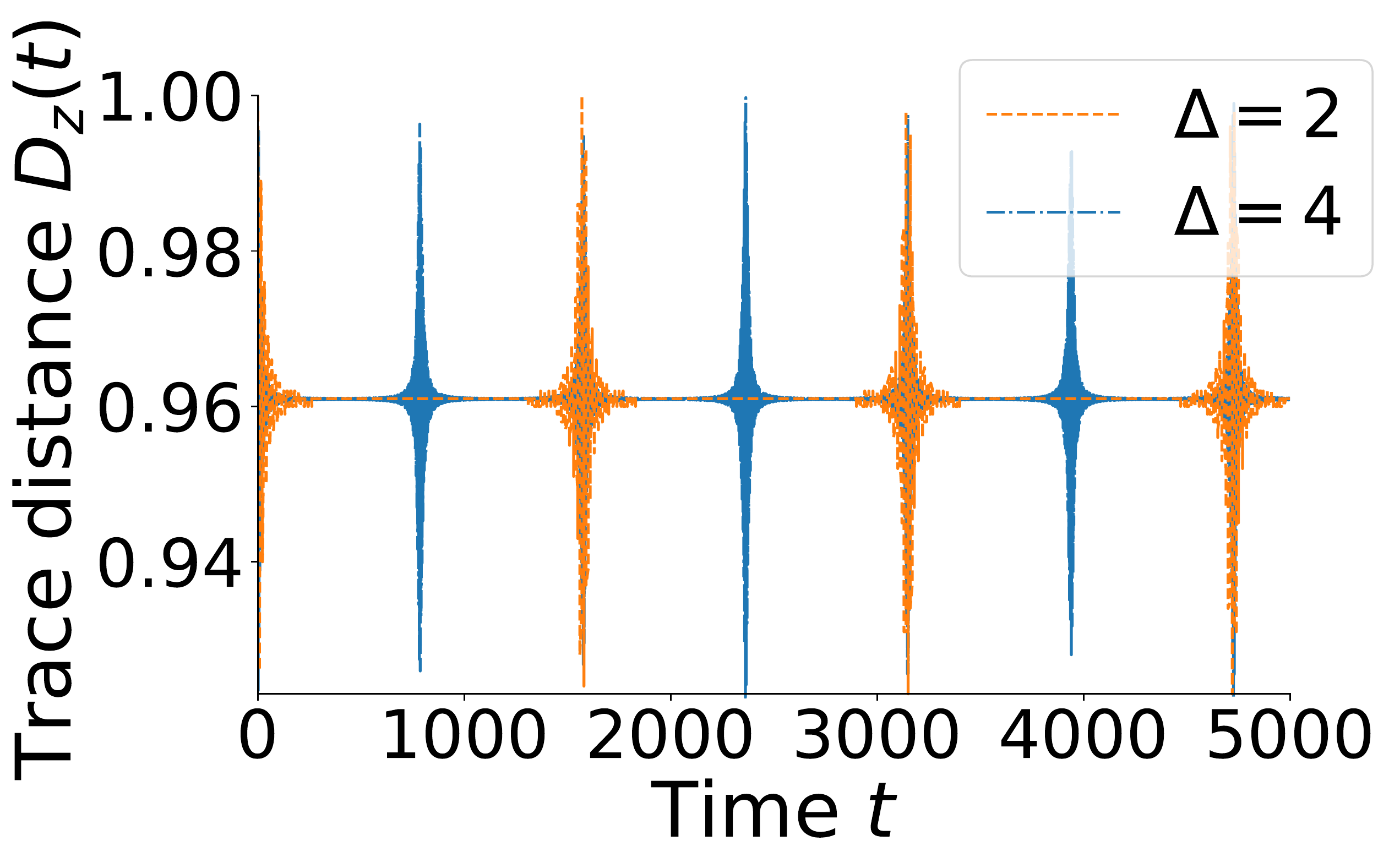}}\subfloat[\centering]{\includegraphics[clip,width=4.2cm]{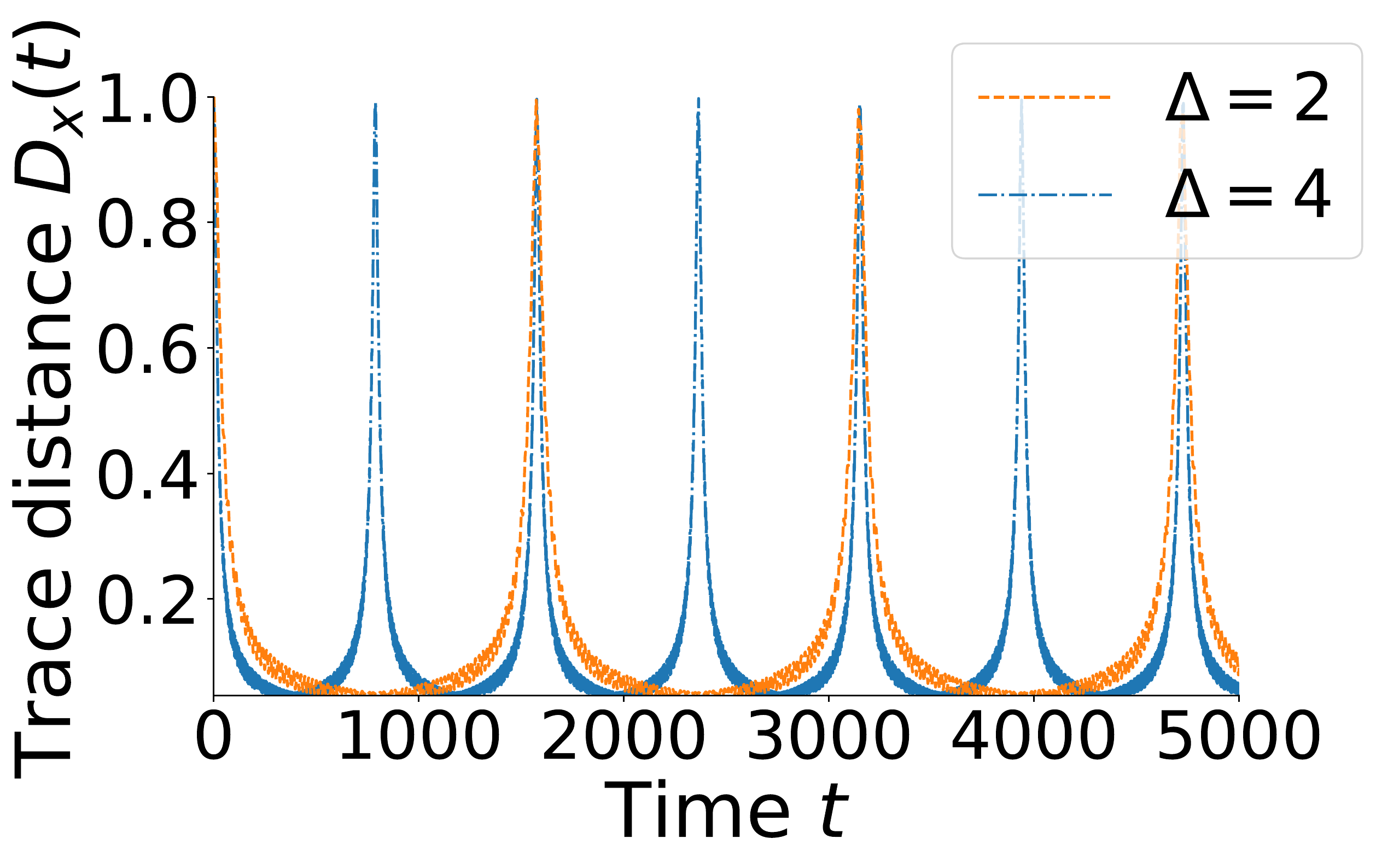}

}

\subfloat[\centering]{\includegraphics[clip,width=4.2cm]{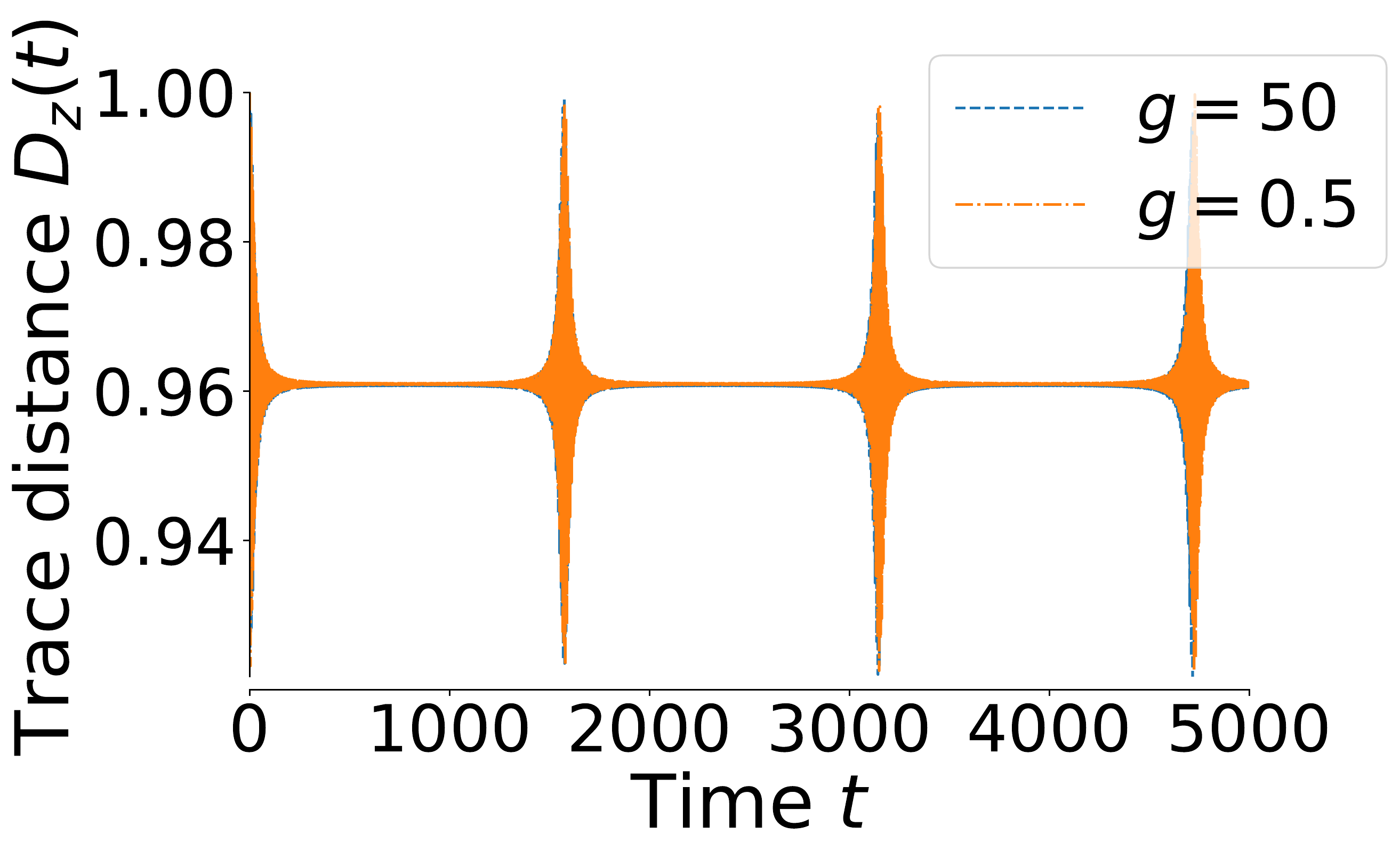}

}\subfloat[\centering]{\includegraphics[clip,width=4.2cm]{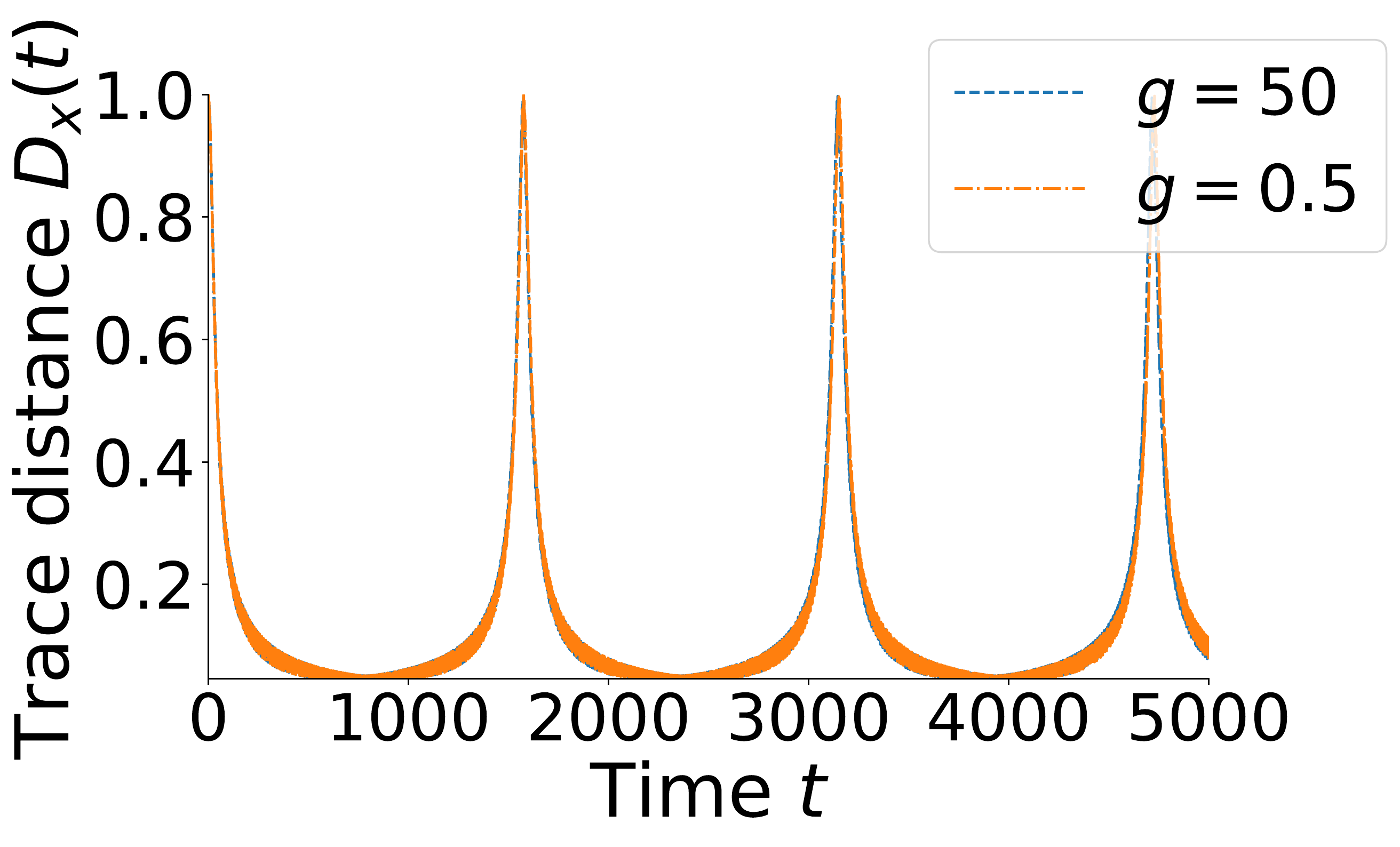}

}

\caption{The behavior of the trace distances {[}(a), (c), (e){]} $D_{z}(t)$
and {[}(b), (d), (f){]} $D_{x}(t)$ with different values of $N$,
$\Delta$, and $g$, presenting collapse-revival patterns with different
periods and amplitudes. {[}(a), (b){]} Impact of $N$ on period, where
a larger $N$ gives a later revival of the information flow, leading
to a weaker non-Markovianity of the central spin; {[}(c), (d){]} influence
of the frequency detuning $\Delta$ on the collapse-revival patterns
with the bath frequency $\omega_{b}$ fixed. It can be seen that the
period of collapse-revival patterns is inverse to $\Delta$; {[}(e),
(f){]} plot of the trace distances with different values of $g$ and
that show that when $g=50$ and $g=0.5$ the two oscillations almost
coincide, implying that the interaction strength $g$ has a negligible
effect on the collapse-revival pattern when it is large. Parameters:
{[}(a), (b){]} $T=10$, $g=1$, $\omega_{s}=3$, and $\omega_{b}=1$;
{[}(c), (d){]} $N=1000$, $T=10$, $g=1$, and $\omega_{b}=1$; {[}(e),
(f){]} $N=1000$, $\omega_{s}=3,$ $\omega_{b}=1$, and $T=10$.}
\label{fig:D2e}
\end{figure}

An interesting case is that if the central qubit is in resonance with
the bath qubits, i.e. $\Delta=0$, or the interaction strength is
zero, $g=0$, the period $T_{cr}$ goes to infinity, which indicates
that the collapse-revival pattern does not exist and only the rapid
oscillation appears in the information backflow. This provides the
condition for the existence of the collapse-revival phenomenon when
$N$ is sufficiently large,
\begin{equation}
g\neq0,\;\Delta\neq0.
\end{equation}
Figure \ref{fig:D2e} shows how the number of bath qubits $N$, the
coupling strength $g$ as well as the system-bath detuning $\Delta$
influence the trace distances $D_{z}(t)$ and $D_{x}(t)$, which verifies
the above analytical results.

\begin{figure}[h]
\subfloat[\centering]{\includegraphics[width=8cm]{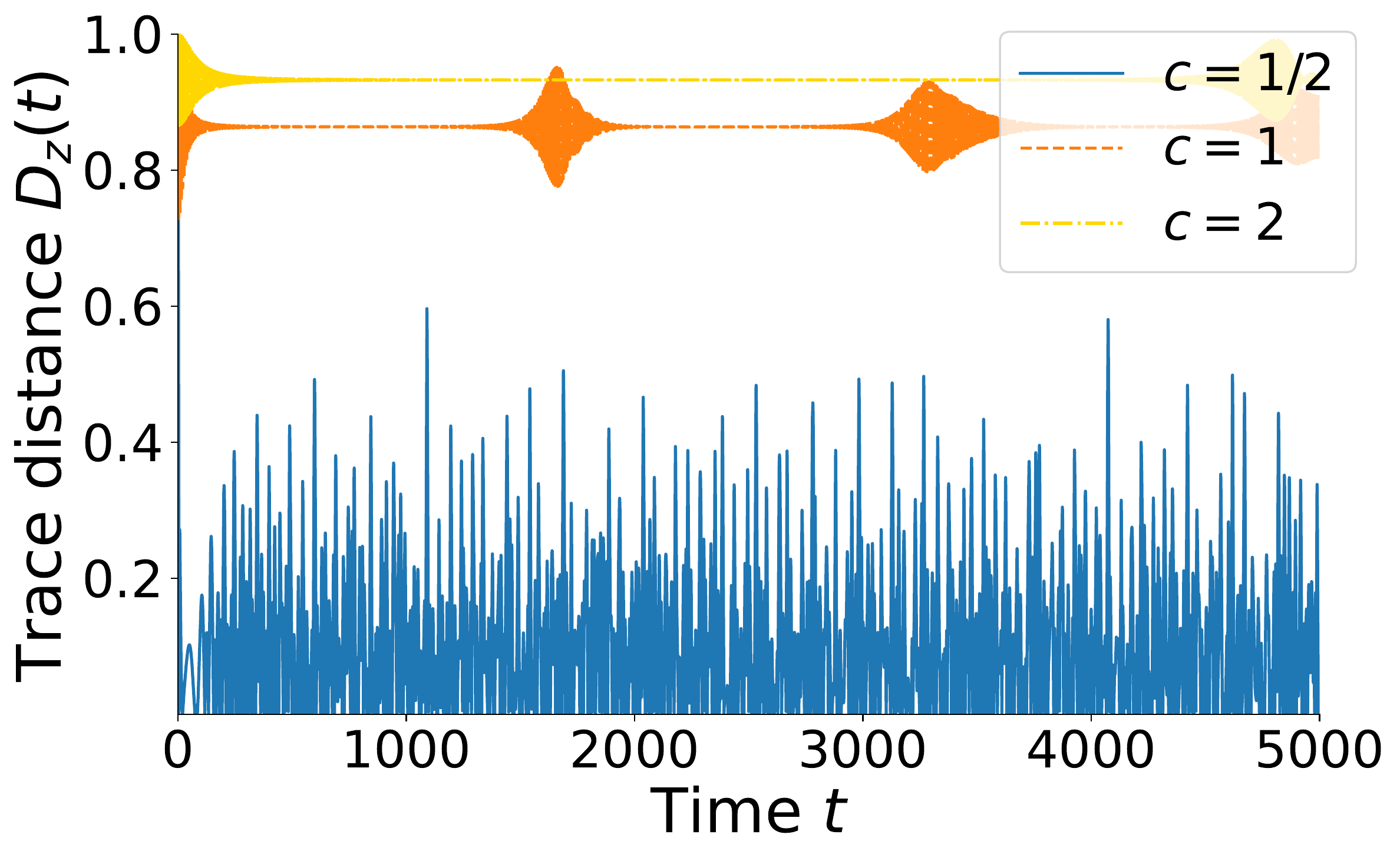} }

\subfloat[\centering]{\includegraphics[clip,width=8cm]{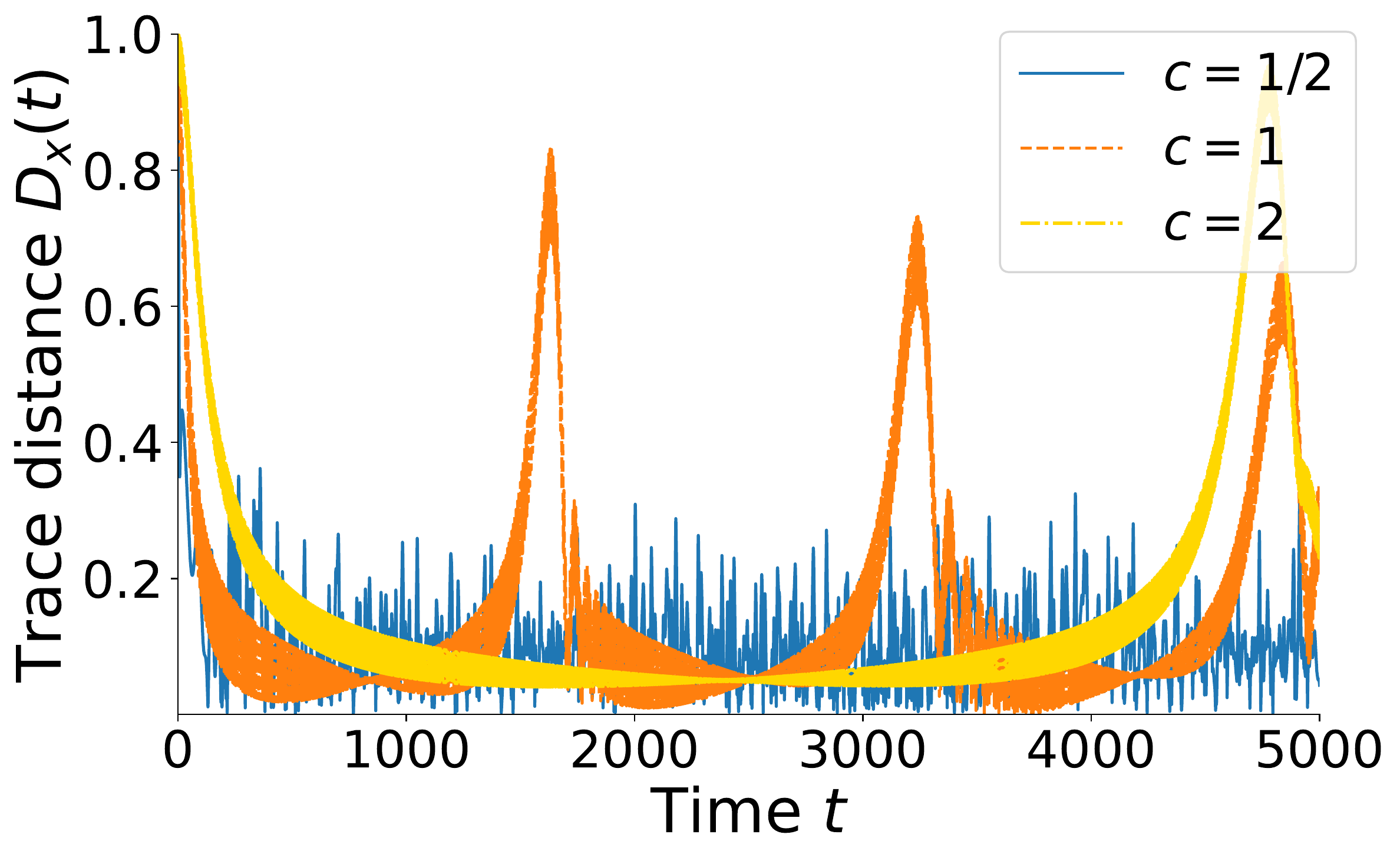}

}

\caption{The behavior of the trace distances (a) $D_{z}(t)$ and (b) $D_{x}(t)$
with different $gN/\Delta$. There is no collapse-revival pattern
and only rapid oscillation occurs when $gN/\Delta=1/2$, while for
$gN/\Delta=1$ and $gN/\Delta=2$, the periods of collapse-revival
patterns are proportional to $2gN/\Delta-1$ for the fixed $g=10^{-3}$.
Parameters: $N=1000$, $\omega_{b}=1$, $T=10$, and $g=10^{-3}$.}
\label{c}
\end{figure}

On the contrary, if $g$ is small so that $gN/\Delta=c$ has magnitude
$O(1)$, the period $T_{cr}$ can be reduced to
\begin{equation}
T_{cr}\doteq\frac{\pi(2c-1)}{2g},\label{tc2}
\end{equation}
Figure \ref{c} describes the collapse-revival patterns for the trace
distances $D_{z}(t)$ and $D_{x}(t)$ with different values of $gN/\Delta$.
In particular, when $gN/\Delta=1/2$, the period $T_{cr}=0$, so there
is no collapse-revival pattern and only the rapid oscillation remains.
For $gN/\Delta=1$ and $gN/\Delta=2$ with a fixed $g$, the periods
of the collapse-revival patterns are proportional to $2gN/\Delta-1$,
in accordance with Eq.\,\eqref{tc2}.

\begin{figure}[h]
\subfloat[\centering]{\includegraphics[width=8cm]{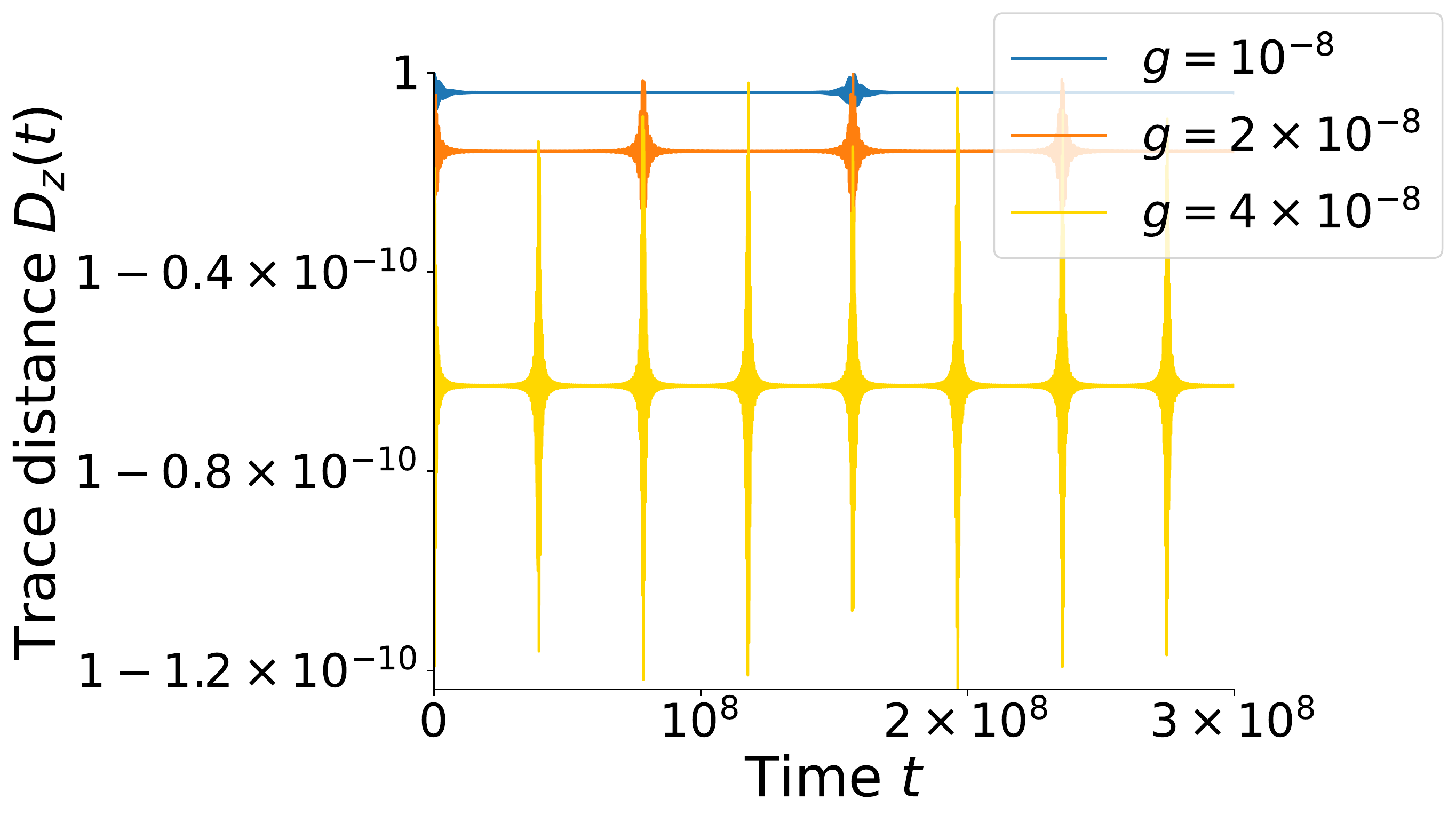}

}

\subfloat[\centering]{\includegraphics[width=7.5cm]{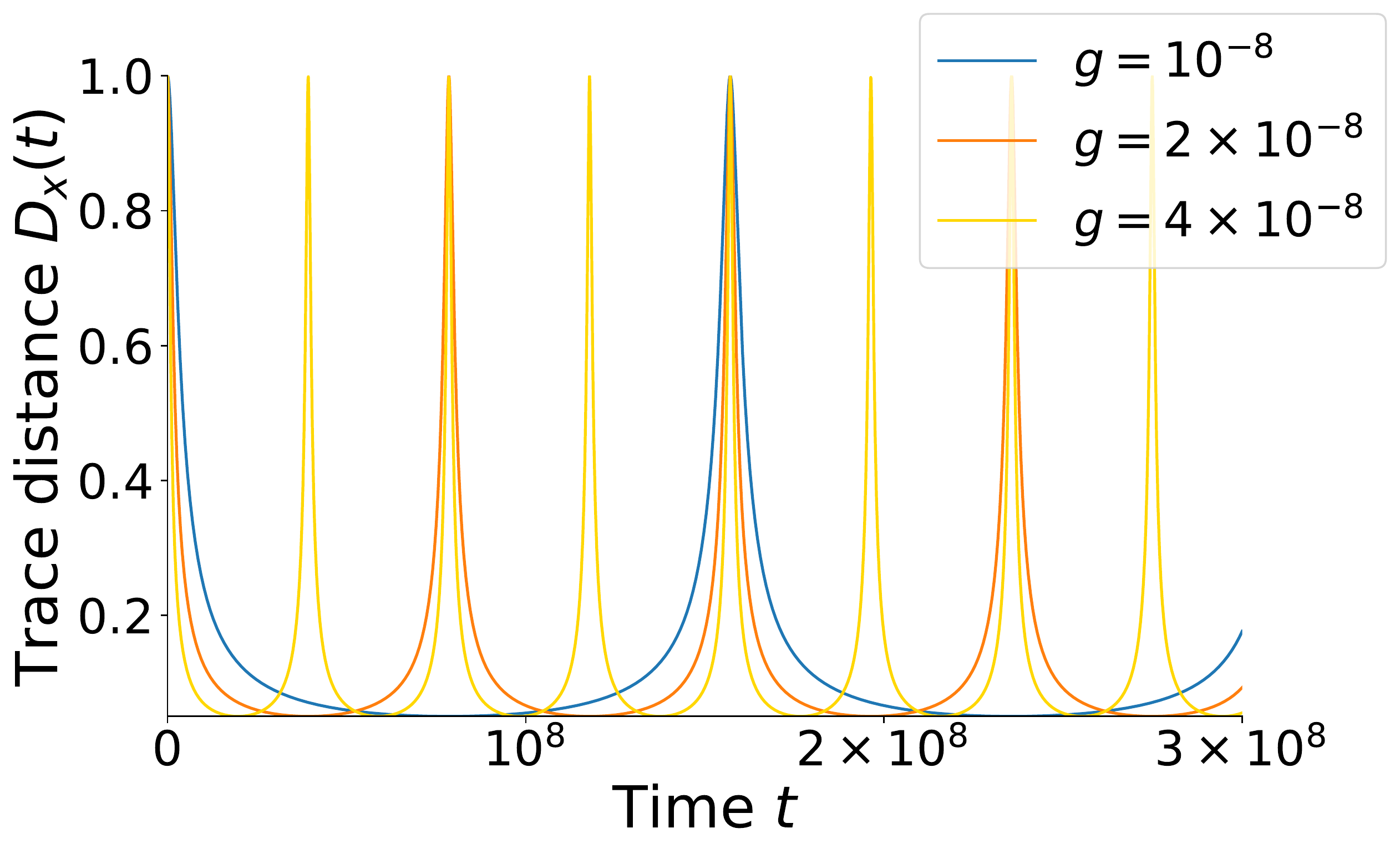}

}

\caption{The behavior of the trace distances (a) $D_{z}(t)$ and (b) $D_{x}(t)$
with sufficiently small values of $g$ so that $gN\ll\Delta$. It
can be seen that the period of the collapse-revival pattern becomes
longer with a smaller $g$, in accordance with the analytical results.
Parameters: $N=1000$, $\omega_{s}=3,$ $\omega_{b}=1$, and $T=10$.}
\label{fig:Dg}
\end{figure}

If $g$ is sufficiently small so that $gN\ll\Delta$, Eq.\,\eqref{eq:tc}
tells that the period is approximately
\begin{equation}
T_{cr}\doteq\frac{\pi}{2g},
\end{equation}
implying only the interaction strength $g$ determines the period
of collapse and revival in this case. This is shown in Fig.\,\ref{fig:Dg}.

\subsection{Relation to the non-Markovianity of the central spin}

In the above figures, the information backflow revives periodically
in the time evolution of the central spin, so the integration over
the increase in the trace distances will diverge. But one can see
that a larger number of bath qubits $N$ leads to a longer collapse
time and a later revival of the information backflow, so the collapse
time of information backflow can characterize the non-Markovianity
of the central spin dynamics in this case.

\begin{figure}[h]
\includegraphics[viewport=50bp 50bp 720bp 585bp,width=9cm]{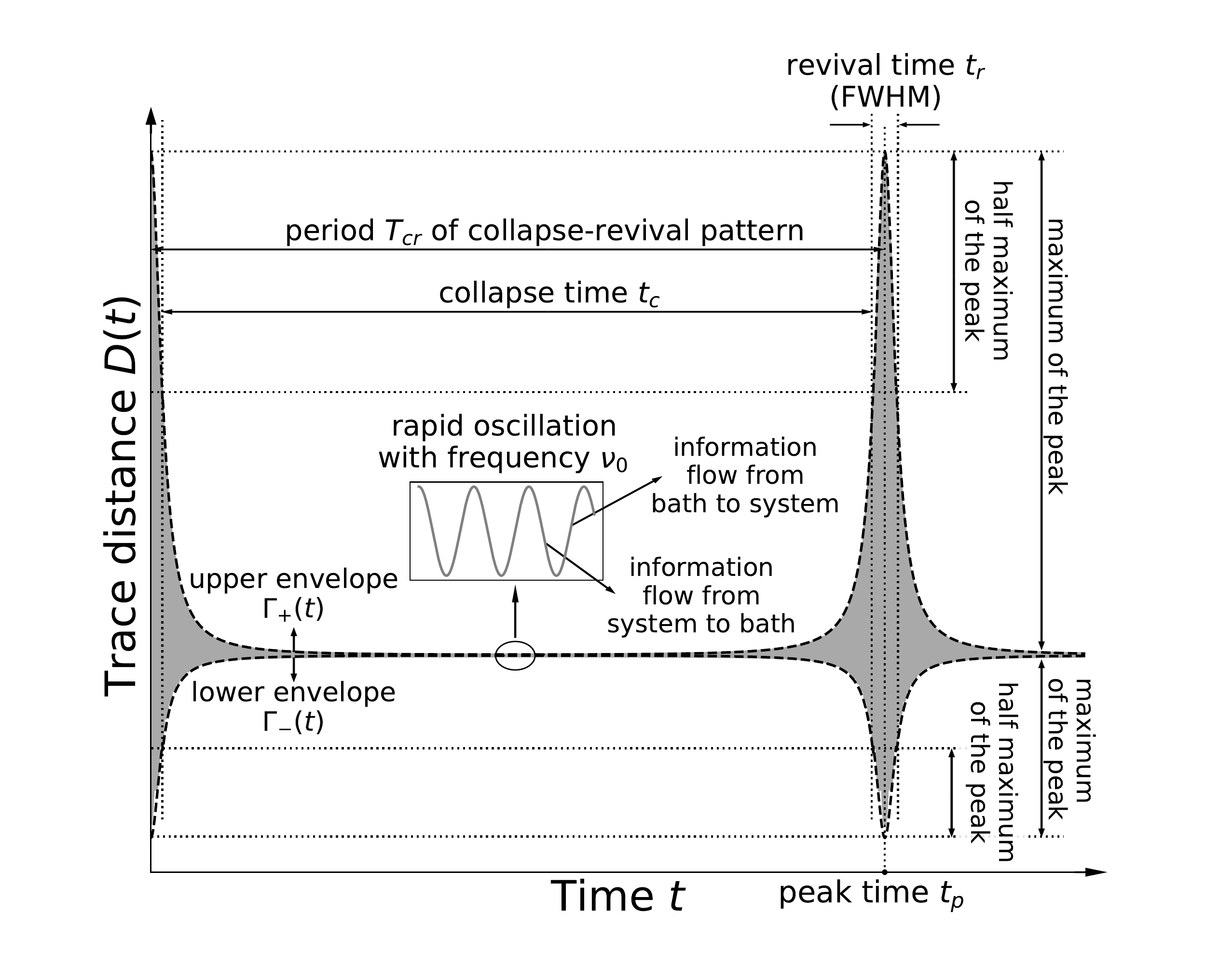}

\caption{The conceptual sketch for the trace distance $D(t)$. The rapid oscillations
of the information flow, the upper and lower envelopes, the various
characteristic time scales and the information loss and backflow processes
are indicated in the figure.}
\label{fig:sketch}
\end{figure}

In order to characterize the non-Markovianity of the central spin
dynamics by the collapse-revival structure, we define the collapse
and revival times of information backflow more precisely. As the trace
distance increases and decreases gradually with time, one cannot find
the exact ``start time'' or ``end time'' of the collapse or revival
of the information flow, so a reasonable way to define the revival
time is the full width at half maximum (FWHM) of a peak in the time
evolution of the trace distance, and the collapse time is the difference
between the period of collapse-revival pattern and the revival time,
or more intuitively the waiting time for the information backflow
to revive. We provide an intuitive illustration of these definitions
in Fig.\,\ref{fig:sketch}.

In detail, the maximum of the envelopes of the trace distance $D(t)$
can be obtained as
\begin{equation}
\begin{aligned}h_{\pm}= & \max_{t}\Gamma_{\pm}(t)-\min_{t}\Gamma_{\pm}(t)\end{aligned}
\end{equation}
where 

\begin{equation}
\begin{aligned}\max_{t}\Gamma_{\pm}(t)= & \frac{1}{2}\sqrt{\alpha_{z}\overline{D_{z}}^{2}+\alpha_{x}\pm\frac{2a_{1}\alpha_{z}\coth\frac{\omega_{b}}{2T}}{N}}\\
\min_{t}\Gamma_{\pm}(t)= & \frac{1}{2}\sqrt{\alpha_{z}\overline{D_{z}}^{2}+\alpha_{x}\tanh^{2}\frac{\omega_{b}}{2T}\pm\frac{2a_{1}\alpha_{z}\tanh\frac{\omega_{b}}{2T}}{N}}
\end{aligned}
\end{equation}
and the signs $\pm$ correspond to the upper and lower envelope lines
of $D(t)$, respectively. Then for a peak of $D(t)$, if $t_{p}$
is the time point that either envelope reaches its maximum, and $t_{p}-\delta,t_{p}+\delta$
are the time points that the envelope reaches the half maximum, i.e.,
\begin{equation}
\Gamma_{\pm}(t_{p}\pm\delta)=\frac{h_{\pm}}{2}=\frac{{\displaystyle 1}}{2}\left[\max_{t}\Gamma_{\pm}(t)+\min_{t}\Gamma_{\pm}(t)\right],\label{eq:gpmtp}
\end{equation}
then the revival time can be defined as
\begin{equation}
t_{r}=2\delta,
\end{equation}
and the collapse time as
\begin{equation}
t_{c}=T_{cr}-2\delta.
\end{equation}
Note that the $\pm$ signs in the term $\Gamma_{\pm}(t_{p}\pm\delta)$
of Eq.\,\eqref{eq:gpmtp} do not change simultaneously. The first
$\pm$ sign determines which envelope line is concerned and the second
$\pm$ sign denotes the two time points that the envelope reaches
the half maximum.

The time points that either envelope reaches its half maximum can
obtained from Eq.\,\eqref{eq:evn} or Eq.\,\eqref{eq:enve}, and
the result turns out to be

\begin{equation}
\begin{aligned}t_{p}= & \frac{2k\pi}{\nu_{cr}},\\
\delta= & \frac{\arccos\left(\cosh\frac{\omega_{b}}{T}-8\alpha_{x}\sinh^{2}\frac{\omega_{b}}{T}w^{-1}\right)}{\nu_{cr}},
\end{aligned}
\end{equation}
where 
\begin{equation}
\begin{aligned}w= & \alpha_{x}\cosh\frac{\omega_{b}}{T}-\alpha_{z}\overline{D_{z}}^{2}(\cosh\frac{\omega_{b}}{T}+1)\\
 & +(\cosh\frac{\omega_{b}}{T}+1)\left(\alpha_{z}\overline{D_{z}}^{2}+\alpha_{x}\right)\sqrt{1-\frac{\alpha_{x}\text{sech}^{2}\frac{\omega_{b}}{2T}}{\alpha_{z}\overline{D_{z}}^{2}+\alpha_{x}}},
\end{aligned}
\end{equation}
and $\,k=0,1,2,3,...$ denotes the $k$th revival. One can immediately
have that the revival time, i.e., the full width at half maximum,
is
\begin{equation}
t_{r}=2\delta=\frac{2\arccos\left(\cosh\frac{\omega_{b}}{T}-8\alpha_{x}\sinh^{2}\frac{\omega_{b}}{T}w^{-1}\right)}{\nu_{cr}},
\end{equation}
and thus the collapse time is
\begin{equation}
t_{c}=T_{cr}-t_{r}=2\frac{\pi-\arccos\left(\cosh\frac{\omega_{b}}{T}-8\alpha_{x}\sinh^{2}\frac{\omega_{b}}{T}w^{-1}\right)}{\nu_{cr}}.
\end{equation}

From these results, one can find that the period of the collapse-revival
pattern, the revival time and the collapse time all increase with
the number of bath spins $N$ as $\nu_{0}$ is linear with $N$ according
to Eq.\,\eqref{eq:v0}, but the ratio between the revival time and
the collapse time keeps constant,
\begin{equation}
\frac{t_{c}}{t_{r}}=\frac{\pi}{\arccos\left(\cosh\frac{\omega_{b}}{T}-8\alpha_{x}\sinh^{2}\frac{\omega_{b}}{T}w^{-1}\right)}-1.
\end{equation}
So, the number of bath spins mainly rescales the collapse-revival
pattern of the trace distance evolution, but does not change the proportion
of the collapse time and the revival time which depends on the initial
state of the system and the bath temperature only. This shows the
way that the dimension of the bath leads the dynamics of the central
spin from non-Markovianity to Markovianity from another perspective,
in addition to the influence of the bath dimension on the amplitude
of the collapse-revival pattern shown in Sec.\,\ref{subsec:Trace-distance z}
and \ref{subsec:Trace-distance x}.

\begin{figure}
\includegraphics[width=8.9cm,height=7cm]{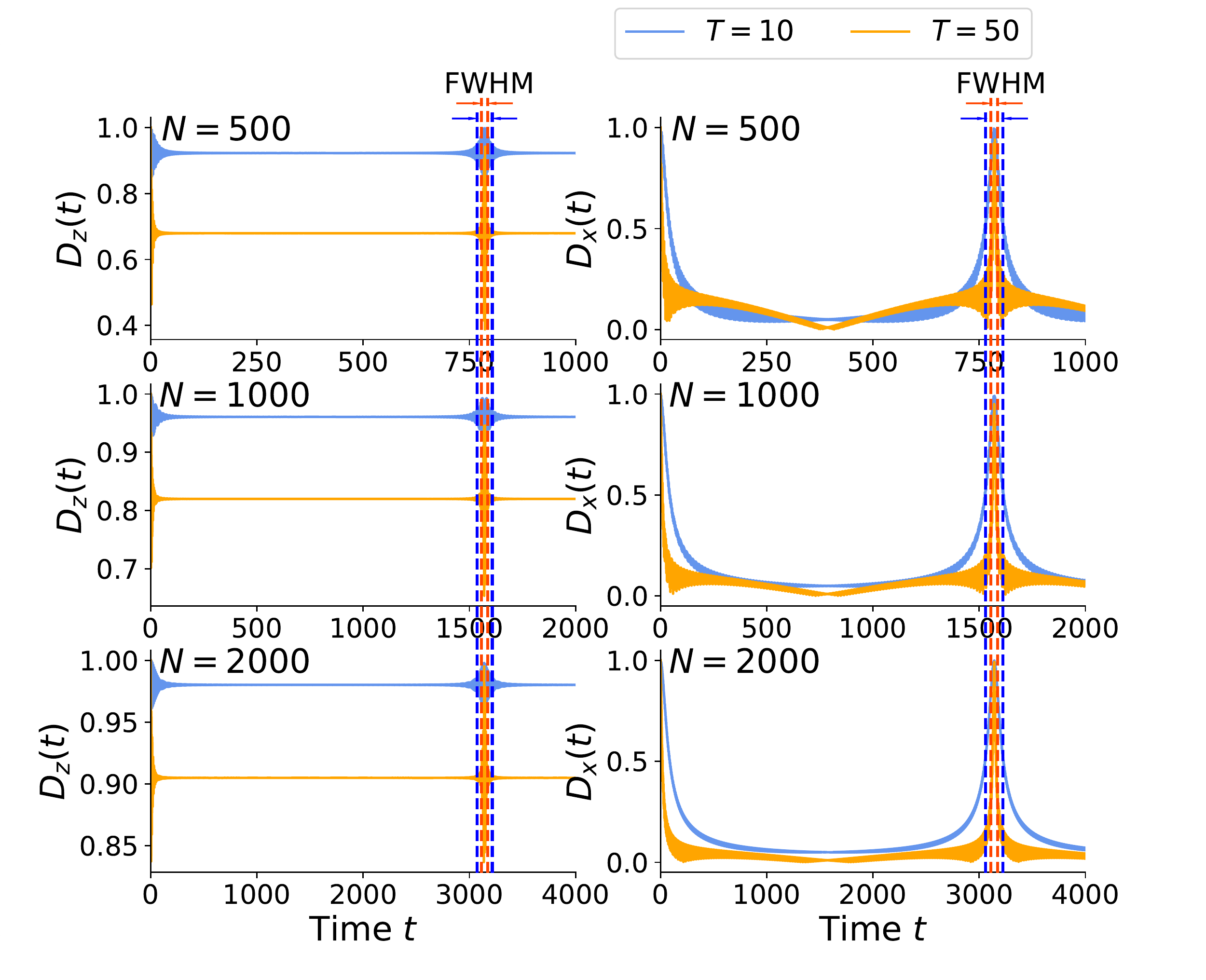}

\caption{The full width at half maximum of a trace distance peak for $D_{z}\left(t\right)$
and $D_{x}\left(t\right)$. The time axes for different panels are
adjusted in proportion to the number of bath spins, $N$, so that
the portions of the collapse times and the revival times can be compared
for different $N$. It shows that the period, the collapse time and
the revival time of the collapse-revival structure are all proportional
to $N$. The vertical dashed lines show that for a given bath temperature,
the FWHM are almost the same for different number of bath spins after
the adjustment of the time axis, so the ratio between the revival
time and the collapse time keeps constant, which is dependent on the
initial state of the central spin and the temperature of bath spins
only. Parameters: $g=1$, $\omega_{s}=3$, and $\omega_{b}=1$.}
\label{fig:hhw}
\end{figure}

Figure \ref{fig:hhw} plots the trace distances $D_{z}(t)$ and $D_{x}(t)$
for different numbers of bath spins, $N$, and different bath temperatures
$T$. The time axes for different $N$ are adjusted in proportion
to $N$, so that one can compare the portion of the collapse time
and revival time for different $N$.

\emph{Remark.} While the periodic patterns of the information backflow
change for different states of the central spin, they share crucial
similarities. The most important one is that the information flow
collapses and reappears when the bath dimension is finite, and the
revival amplitude does not reduce over time, implying the information
backflow can occur periodically for an arbitrary evolution time. As
the revival of the information backflow indicates a violation of the
CPTP divisibility, the more frequent revivals of the information backflow
imply a stronger non-Markovianity of the system dynamics. Therefore,
the period and the collapse time of the collapse-revival pattern may
serve as a characterization of the non-Markovianity in this case,
while the integration over the information backflow may diverge. As
is shown in this section, both the period and the collapse time of
the collapse-revival pattern are proportional to the number of bath
spins when the number of bath spins is sufficiently large, so a larger
bath dimension leads to later revivals of information backflow. When
the number of the bath spins goes to infinity, the period and the
collapse time will become infinitely long, so the revival of the information
backflow will actually never occur in this limit. This tells the role
of the bath dimension in the Markovianity of quantum dynamics and
shows how the transition of quantum dynamics from non-Markovian to
Markovian occurs when the bath grows from finite dimension to infinite
dimension.

\section{Conclusion and outlook\label{sec:Conclusion-and-outlook}}

In this work we consider a simple but nontrivial isotropic central
spin model to analyze the influence of bath dimension on the non-Markovianity
of the system dynamics. We obtain the dynamics of the central spin
with the bath spins in a symmetric thermal equilibrium state initially,
and compute the trace distance for different pairs of initial states
of the central spin to study the non-Markovianity of the system dynamics.

We mainly work in the regime where the number of bath spins, $N$,
is sufficiently large compared to $T/\omega_{b}$ but still finite.
In this case, approximate results are obtained for the trace distances
between arbitrary system states. The results show that oscillations
with dramatically different frequencies appear in the trace distances,
which leads to the collapse and revival phenomenon in the time evolution
of the trace distances. We obtain the conditions for the existence
of the collapse-revival phenomenon, analyze the roles of different
physical parameters such as the interaction strength, the system-bath
detuning, etc. in the trace distances in detail, and derive typical
characteristic time scales of the trace distance, including the period,
the collapse time, and the revival time of the information backflow.
These results show how the collapse-revival pattern changes with an
increasing number of bath qubits, and reveal the effect of bath dimension
in the transition of non-Markovian quantum dynamics to Markovian quantum
dynamics.

The results show that the collapse and revival of information backflow
does not recede with time and occurs periodically in the current model.
A larger number of bath spins or weaker system-bath interaction will
give a later and less frequent revival of the information backflow,
and the information backflow will finally vanish when the number of
bath spins goes to infinity or the system-bath interaction goes to
zero. This shows how the transition of the Markovianity of the central
spin dynamics occurs in the limit of large number of bath spins.

We hope this work can provide a new perspective on the non-Markovianity
of quantum dynamics, particularly in the presence of a large but finite-dimensional
environment, and a useful approach to the characterization of non-Markovianity
for this case.
\begin{acknowledgments}
The authors acknowledge the helpful discussions with Yutong Huang
and Junyan Li. This work is supported by the National Natural Science
Foundation of China (Grant No. 12075323).
\end{acknowledgments}

\appendix
\onecolumngrid

\section{EVOLUTION OF THE SYSTEM\label{sec:EVOLUTION-OF-THE}}

\subsection{Eigenvalues and eigenstates of total Hamiltonian\label{subsec:Eigenvalues-and-eigenstates}}

A symmetric thermal state can be represented by a superposition of
Dicke states, which inspires us to consider the evolution of a system-bath
joint state $\lvert\psi_{sb}^{(M)}\left(0\right)\rangle=\left(\zeta_{0}\lvert0\rangle+\zeta_{1}\lvert1\rangle\right)\otimes\lvert J,M\rangle$
under the total Hamiltonian
\begin{equation}
H_{tot}=\frac{\omega_{s}}{2}\sigma_{z}^{(s)}+\omega_{b}S_{z}+2g\left(\sigma_{+}^{(s)}S_{-}+\sigma_{-}^{(s)}S_{+}+\sigma_{z}^{(s)}S_{z}\right).
\end{equation}
The initial joint state of the system and bath can be written as
\begin{equation}
\rho_{sb}\left(0\right)=\sum_{M=-J}^{J}\frac{e^{-\frac{M\omega_{b}}{T}}}{Q}\lvert\psi_{sb}^{(M)}\left(0\right)\rangle\langle\psi_{sb}^{(M)}\left(0\right)\rvert,\,\,Q=\frac{e^{\frac{(J+1)\omega_{b}}{T}}-e^{-\frac{J\omega_{b}}{T}}}{e^{\frac{\omega_{b}}{T}}-1}.
\end{equation}
After the time evolution $U\left(t\right)=e^{-iH_{tot}t}$, the joint
state evolves into 
\begin{equation}
\rho_{sb}\left(t\right)=U\left(t\right)\rho_{sb}\left(0\right)U\left(t\right)^{\dagger}=\sum_{M=-J}^{J}\frac{e^{-\frac{M\omega_{b}}{T}}}{Q}\lvert\psi_{sb}^{(M)}\left(t\right)\rangle\langle\psi_{sb}^{(M)}\left(t\right)\rvert,
\end{equation}
where $\lvert\psi_{sb}^{(M)}\left(t\right)\rangle=e^{-iH_{tot}t}\lvert\psi_{sb}^{(M)}\left(0\right)\rangle$.

Note that the subspaces spanned by the pairs of states $\left\{ \lvert0\rangle\otimes\lvert J,M\rangle,\lvert1\rangle\otimes\lvert J,M-1\rangle\right\} $,
$-J+1\leq M\leq J$, are invariant under the total Hamiltonian. One
can find the reduced Hamiltonian in the each subspace to be
\begin{equation}
H_{M}=\left[\begin{array}{cc}
\frac{\omega_{s}}{2}-M\left(\omega_{b}+2g\right) & 2g\sqrt{\left(J-M+1\right)\left(J+M\right)}\\
2g\sqrt{\left(J-M+1\right)\left(J+M\right)} & -\frac{\omega_{s}}{2}-(M-1)\left(\omega_{b}-2g\right)
\end{array}\right].
\end{equation}
Then the eigenvalues of $H_{M}$ can be obtained as
\begin{equation}
\lambda_{M,\pm}=E_{M}\pm F_{M},
\end{equation}
where $E_{M}$ and $F_{M}$ are functions dependent on $M$,

\begin{equation}
E_{M}=-g+\frac{2M-1}{2}\omega_{b},\,\,F_{M}=\sqrt{G_{M}^{2}+4\left(J-M+1\right)\left(J+M\right)g^{2}}.
\end{equation}
Here $G_{M}=\left(2M-1\right)g+\frac{\Delta}{2},$and $\Delta=\omega_{s}-\omega_{b}$
is the frequency detuning. The eigenstates of $H_{M}$ can also be
obtained,
\begin{equation}
\lvert\Phi\rangle_{M,\pm}=c_{M,\pm}\lvert0\rangle\lvert J,M\rangle+d_{M,\pm}\lvert1\rangle\lvert J,M-1\rangle,\label{eigenstate-1}
\end{equation}
corresponding to the eigenvalues $\lambda_{M,\pm}$ respectively,
where
\begin{equation}
c_{M,\pm}=\pm{\rm sgn}(g)\sqrt{\frac{1}{2}\left(1\mp\frac{G_{M}}{F_{M}}\right)},\;d_{M,\pm}=\sqrt{\frac{1}{2}\left(1\pm\frac{G_{M}}{F_{M}}\right)}.
\end{equation}

According to Eq.\,\eqref{eigenstate-1}, one can write the states
$\lvert0\rangle\otimes\lvert J,M\rangle$ and $\lvert1\rangle\otimes\lvert J,M-1\rangle$
as a superposition of the eigenstates $\lvert\Phi\rangle_{M,\pm}$,

\begin{equation}
\begin{aligned}\lvert0\rangle\lvert J,M\rangle= & \frac{d_{M,-}}{K_{M}}\lvert\Phi\rangle_{M,+}-\frac{d_{M,+}}{K_{M}}\lvert\Phi\rangle_{M,-},\\
\lvert1\rangle\lvert J,M-1\rangle= & -\frac{c_{M,-}}{K_{M}}\lvert\Phi\rangle_{M,+}+\frac{c_{M,+}}{K_{M}}\lvert\Phi\rangle_{M,-},
\end{aligned}
\label{eq:01j}
\end{equation}
where $K_{M}=c_{M,+}d_{M,-}-c_{M,-}d_{M,+}$ and it can be verified
that $K_{M}={\rm sgn}(g)$.

Two additional eigenstates, $\lvert\Phi\rangle_{-J}=\lvert0\rangle\otimes\lvert J,-J\rangle$
and $\lvert\Phi\rangle_{J+1}=\lvert1\rangle\otimes\lvert J,J\rangle$
are also contained in the above invariant subspaces with $M=-J$ and
$M=J+1$ respectively, and corresponding eigenvalues are
\begin{equation}
\lambda_{-J}=E_{M}-{\rm sgn}(G_{M})G_{M},\;\lambda_{J+1}=E_{M}+{\rm sgn}(G_{M})G_{M}.
\end{equation}
Note that there is one non-physical eigenstate in each of those two
invariant subspaces, $\lvert1\rangle\otimes\lvert J,-J-1\rangle$
for $M=-J$ and $\lvert0\rangle\otimes\lvert J,M\rangle$ for $M=J+1$.
However, these two non-physical eigenstates will not affect the validity
of Eq.\,\eqref{eq:01j} with $M=-J,J+1$, since they vanish in Eq.\,\eqref{eq:01j}
when $M$ takes $-J$ or $J+1$,
\begin{equation}
\begin{aligned}\lvert0\rangle\otimes\lvert J,-J\rangle= & (d_{J,-}c_{J,+}-d_{J,+}c_{J,-})\lvert0\rangle\otimes\lvert J,-J\rangle/K_{J}=\lvert0\rangle\otimes\lvert J,-J\rangle,\\
\lvert1\rangle\otimes\lvert J,J\rangle= & (-c_{J+1,-}d_{J+1,+}+c_{J+1,+}d_{J+1,-})\lvert1\rangle\otimes\lvert J,J\rangle/K_{J+1}=\lvert1\rangle\otimes\lvert J,J\rangle.
\end{aligned}
\end{equation}
So the general eigenstate expression \eqref{eigenstate-1} can also
work for the two additional eigenstates.

\subsection{Joint evolution of the system and bath}

The eigenvalues and the eigenstates of $H_{tot}$ lead to the derivation
of the exact reduced dynamics of the central spin. The evolved state
can be decomposed into the eigenstates of $H_{tot}$,

\begin{equation}
\lvert\psi_{sb}^{(M)}\left(0\right)\rangle=e^{-iH_{tot}t}\left(\zeta_{0}\lvert0\rangle+\zeta_{1}\lvert1\rangle\right)\lvert J,M\rangle=\sum_{M^{\prime}=-J}^{J+1}\sum_{\pm}D_{M^{\prime},\pm}e^{-i\lambda_{M^{\prime},\pm}t}\lvert\Phi\rangle_{M^{\prime},\pm},
\end{equation}
where 
\begin{equation}
D_{M^{\prime},\pm}=\left(\pm\zeta_{0}d_{M^{\prime},\mp}\delta_{M,M^{\prime}}\mp\zeta_{1}c_{M^{\prime},\mp}\delta_{M,M^{\prime}-1}\right)/K_{M}.
\end{equation}
The reduced evolution of the central spin can be obtained as
\begin{equation}
\rho_{s}\left(t\right)=\tr[b]\rho_{sb}\left(t\right)=\sum_{M,M^{\prime\prime}=-J}^{J}\frac{e^{-\frac{M\omega_{b}}{T}}}{Q}\langle J,M^{\prime\prime}\lvert\psi_{SM}\left(t\right)\rangle\langle\psi_{SM}\left(t\right)\rvert J,M^{\prime\prime}\rangle.
\end{equation}

To facilitate the computation, $\rho_{s}\left(t\right)$ can be written
in the matrix form

\begin{equation}
\rho_{s}\left(t\right)=\left[\begin{array}{cc}
\rho_{00}(t) & \rho_{01}(t)\\
\rho_{10}(t) & \rho_{11}(t)
\end{array}\right],
\end{equation}
where

\begin{align}
\rho_{00}(t)= & \sum_{M,M^{\prime\prime}=-J}^{J}\frac{e^{-\frac{M\omega_{b}}{T}}}{Q}\left(\sum_{\pm}D_{M^{\prime\prime},\pm}e^{-i\lambda_{M^{\prime\prime},\pm}t}c_{M^{\prime\prime},\pm}\right)\left(\sum_{\pm}D_{M^{\prime\prime},\pm}^{*}e^{i\lambda_{M^{\prime\prime},\pm}t}c_{M^{\prime\prime},\pm}\right)\\
= & \left|\zeta_{0}\right|^{2}-\left(\left|\zeta_{0}\right|^{2}-e^{\frac{\omega_{b}}{T}}\left|\zeta_{1}\right|^{2}\right)\sum_{M=-J}^{J}\frac{e^{-\frac{M\omega_{b}}{T}}}{Q}\frac{1}{2}\left(1-\frac{G_{M}^{2}}{F_{M}^{2}}\right)\left[1-\cos\left(2F_{M}t\right)\right],\nonumber 
\end{align}
\begin{align}
\rho_{11}(t)= & \sum_{M,M^{\prime\prime}=-J}^{J}\frac{e^{-\frac{M\omega_{b}}{T}}}{Q}\left(\sum_{\pm}D_{M^{\prime\prime}+1,\pm}e^{-i\lambda_{M^{\prime\prime}+1,\pm}t}d_{M^{\prime\prime}+1,\pm}\right)\left(\sum_{\pm}D_{M^{\prime\prime}+1,\pm}^{*}e^{i\lambda_{M^{\prime\prime}+1,\pm}t}d_{M^{\prime\prime}+1,\pm}\right)\\
= & \left|\zeta_{1}\right|^{2}+\left(\left|\zeta_{0}\right|^{2}-e^{\frac{\omega_{b}}{T}}\left|\zeta_{1}\right|^{2}\right)\sum_{M=-J}^{J}\frac{e^{-\frac{M\omega_{b}}{T}}}{Q}\frac{1}{2}\left(1-\frac{G_{M}^{2}}{F_{M}^{2}}\right)\left[1-\cos\left(2F_{M}t\right)\right],\nonumber 
\end{align}
\begin{align}
\rho_{01}(t)= & \sum_{M,M^{\prime\prime}=-J}^{J}\frac{e^{-\frac{M\omega_{b}}{T}}}{Q}\left(\sum_{\pm}D_{M^{\prime\prime},\pm}e^{-i\lambda_{M^{\prime\prime},\pm}t}c_{M^{\prime\prime},\pm}\right)\left(\sum_{\pm}D_{M^{\prime\prime}+1,\pm}^{*}e^{i\lambda_{M^{\prime\prime}+1,\pm}t}d_{M^{\prime\prime}+1,\pm}\right)\\
= & \zeta_{0}\zeta_{1}^{*}\sum_{M=-J}^{J}\frac{e^{-\frac{M\omega_{b}}{T}}}{Q}\left[\cos\left(\omega_{b}t\right)+i\sin\left(\omega_{b}t\right)\right]\left[\cos\left(F_{M}t\right)+i\frac{G_{M}}{F_{M}}\sin\left(F_{M}t\right)\right]\left[\cos\left(F_{M+1}t\right)+i\frac{G_{M+1}}{F_{M+1}}\sin\left(F_{M+1}t\right)\right],\nonumber 
\end{align}
\begin{align}
\rho_{10}(t)= & \rho_{01}^{*}(t)\\
= & \zeta_{0}^{*}\zeta_{1}\sum_{M=-J}^{J}\frac{e^{-\frac{M\omega_{b}}{T}}}{Q}\left[\cos\left(\omega_{b}t\right)-i\sin\left(\omega_{b}t\right)\right]\left[\cos\left(F_{M}t\right)-i\frac{G_{M}}{F_{M}}\sin\left(F_{M}t\right)\right]\left[\cos\left(F_{M+1}t\right)-i\frac{G_{M+1}}{F_{M+1}}\sin\left(F_{M+1}t\right)\right].\nonumber 
\end{align}

If we represent the final density matrix of the central spin by a
Bloch vector $\boldsymbol{v}(t)=[x(t),y(t),z(t)]$, then $\boldsymbol{v}(t)$
can be worked out as

\begin{align}
x(t)= & \rho_{01}(t)+\rho_{10}(t)\nonumber \\
= & \left(\zeta_{0}\zeta_{1}^{*}+\zeta_{0}^{*}\zeta_{1}\right)X_{1}\left(t\right)+i\left(\zeta_{0}\zeta_{1}^{*}-\zeta_{0}^{*}\zeta_{1}\right)X_{2}\left(t\right)\nonumber \\
= & x_{0}X_{1}\left(t\right)+y_{0}X_{2}\left(t\right),\nonumber \\
y(t)= & i\left(\rho_{01}(t)-\rho_{10}(t)\right)\nonumber \\
= & i\left(\zeta_{0}\zeta_{1}^{*}-\zeta_{0}^{*}\zeta_{1}\right)X_{1}\left(t\right)-\left(\zeta_{0}\zeta_{1}^{*}+\zeta_{0}^{*}\zeta_{1}\right)X_{2}\left(t\right)\label{Bloch of final}\\
= & y_{0}X_{1}\left(t\right)-x_{0}X_{2}\left(t\right),\nonumber \\
z(t)= & \rho_{00}(t)-\rho_{11}(t)\nonumber \\
= & z_{0}-z_{0}\left(1+e^{\frac{\omega_{b}}{T}}\right)Z(t)-\left(1-e^{\frac{\omega_{b}}{T}}\right)Z(t)\nonumber \\
= & z_{0}Z_{1}\left(t\right)+Z_{2}\left(t\right),\nonumber 
\end{align}
where $X_{1}\left(t\right)$, $X_{2}\left(t\right)$ , $Z_{1}\left(t\right)$
and $Z_{2}\left(t\right)$ are
\begin{align}
X_{1}\left(t\right)= & \sum_{M=-J}^{J}\frac{e^{-\frac{M\omega_{b}}{T}}}{Q}\left[\sin\left(\omega_{b}t\right)A_{M}\left(t\right)+\cos\left(\omega_{b}t\right)B_{M}\left(t\right)\right],\label{eq:x1t}\\
X_{2}\left(t\right)= & \sum_{M=-J}^{J}\frac{e^{-\frac{M\omega_{b}}{T}}}{Q}\left[\cos\left(\omega_{b}t\right)A_{M}\left(t\right)-\sin\left(\omega_{b}t\right)B_{M}\left(t\right)\right],\label{eq:x2t}\\
Z_{1}\left(t\right)= & 1-\left(1+e^{\frac{\omega_{b}}{T}}\right)\sum_{M=-J}^{J}\frac{e^{-\frac{M\omega_{b}}{T}}}{Q}C_{M}\left(t\right),\label{eq:z1t}\\
Z_{2}\left(t\right)= & \left(e^{\frac{\omega_{b}}{T}}-1\right)\sum_{M=-J}^{J}\frac{e^{-\frac{M\omega_{b}}{T}}}{Q}C_{M}\left(t\right).\label{eq:z2t}
\end{align}
In the above equations, $A_{M}\left(t\right)$, $B_{M}\left(t\right)$
and $C_{M}\left(t\right)$ are defined as
\begin{align}
A_{M}\left(t\right)= & -\frac{G_{M+1}}{F_{M+1}}\cos\left(F_{M}t\right)\sin\left(F_{M+1}t\right)-\frac{G_{M}}{F_{M}}\sin\left(F_{M}t\right)\cos\left(F_{M+1}t\right),\nonumber \\
B_{M}\left(t\right)= & \cos\left(F_{M}t\right)\cos\left(F_{M+1}t\right)-\frac{G_{M}}{F_{M}}\frac{G_{M+1}}{F_{M+1}}\sin\left(F_{M}t\right)\sin\left(F_{M+1}t\right),\label{eq:ambmcm}\\
C_{M}\left(t\right)= & \frac{1}{2}\left(1-\frac{G_{M}^{2}}{F_{M}^{2}}\right)\left[1-\cos\left(2F_{M}t\right)\right].\nonumber 
\end{align}

\section{the derivation of $Z_{1}\left(t\right)$\label{sec:the-approximation-of}}

The approximation of $Z_{1}\left(t\right)$ can be obtained by replacing
$M$ with $\xi=(M+J)/N$. We keep the terms up to $O(1/N^{2})$, and
it turns out to be

\begin{equation}
\begin{aligned}Z_{1}\left(t\right)= & 1-\left(1+e^{\frac{\omega_{b}}{T}}\right)\sum_{M=-J}^{J}\frac{e^{-\frac{M\omega_{b}}{T}}}{Q}\frac{1}{2}\left(1-\frac{G_{M}^{2}}{F_{M}^{2}}\right)\left[1-\cos\left(2F_{M}t\right)\right]\\
= & 1-\left(1+e^{\frac{\omega_{b}}{T}}\right)\frac{e^{\frac{N\omega_{b}}{2T}}}{Q}\sum_{\xi}e^{-\frac{\xi N\omega_{b}}{T}}\sum_{j}\left(a_{j}\xi^{j}\right)\left[1-\cos\left(\sum_{k}\nu_{k}\xi^{k}t\right)\right]\\
\simeq & 1-\left(1+e^{\frac{\omega_{b}}{T}}\right)\frac{e^{\frac{N\omega_{b}}{2T}}}{Q}\sum_{\xi}e^{-\frac{\xi N\omega_{b}}{T}}\left(a_{0}+a_{1}\xi+a_{2}\xi^{2}\right)\left[1-\cos\left(\nu_{0}t+\nu_{1}\xi t\right)\right],
\end{aligned}
\label{eq:z1}
\end{equation}
where $a_{0}=0$, $a_{1}={\displaystyle \frac{8g^{2}N(N+1)}{\left[2(N+1)g-\Delta\right]^{2}}}$,
$a_{2}={\displaystyle -\frac{8g^{2}N^{2}\left[2(N+1)g+\Delta\right]^{2}}{\left[2(N+1)g-\Delta\right]^{4}}}$,
$\nu_{0}=\left|2(N+1)g-\Delta\right|$ and  $\nu_{1}=\frac{4Ng\Delta}{\nu_{0}}$.
Here the power series for the amplitude and the phase are valid for
$g{\displaystyle <\frac{\left(3N+1\right)\Delta-2\sqrt{\Delta^{2}N(2N+1)}}{2(N+1)^{2}}}$
and ${\displaystyle g>\frac{\left(3N+1\right)\Delta+2\sqrt{\Delta^{2}N(2N+1)}}{2(N+1)^{2}}}$
to guarantee the convergence of the Taylor series. The summation of
$\xi$ runs from $0$ to $1$ with a step size $1/N$ and can be obtained
analytically,

\begin{equation}
\begin{aligned}\sum_{\xi}e^{\frac{-\xi N\omega_{b}}{T}}\left[1-\cos\left(\nu_{0}t+\nu_{1}\xi t\right)\right]\simeq & e^{\frac{\omega_{b}}{2T}}\left[\frac{1}{\sqrt{2P_{-}\left(\frac{\nu_{1}t}{N}\right)}}\cos\left(\nu_{0}t-\phi_{0}\right)+\frac{\text{csch}\frac{\omega_{b}}{2T}}{2}\right],\\
\sum_{\xi}e^{\frac{-\xi N\omega_{b}}{T}}\xi\left[1-\cos\left(\nu_{0}t+\nu_{1}\xi t\right)\right]\simeq & \frac{1}{2NP_{-}\left(\frac{\nu_{1}t}{N}\right)}\cos\left(\nu_{0}t-\phi_{1}\right)+\frac{\text{csch}^{2}\frac{\omega_{b}}{2T}}{4N},\\
\sum_{\xi}e^{\frac{-\xi N\omega_{b}}{T}}\xi^{2}\left[1-\cos\left(\nu_{0}t+\nu_{1}\xi t\right)\right]\simeq & \frac{1}{2N^{2}}\sqrt{\frac{P_{+}\left(\frac{\nu_{1}t}{N}\right)}{P_{-}^{3}\left(\frac{\nu_{1}t}{N}\right)}}\cos\left(\nu_{0}t-\phi_{2}\right)+\frac{\sinh\frac{\omega_{b}}{T}\text{csch}^{4}\frac{\omega_{b}}{2T}}{8N^{2}},
\end{aligned}
\label{eq:xik}
\end{equation}
where the function
\begin{equation}
P_{\pm}\left(x\right)=\cosh\frac{\omega_{b}}{T}\pm\cos x
\end{equation}
will be critical to the characteristic time scales of the collapse-revival
pattern, and the sign $\simeq$ denotes that only the terms $e^{\frac{\omega_{b}}{T}O(N)}$
remain and the terms $e^{\frac{\omega_{b}}{T}O(1)}$ are neglected
since $N$ is large. The phases $\phi_{0},\phi_{1},$and $\phi_{2}$
are

\begin{align}
\phi_{0} & =\arctan\frac{\sin\frac{\text{\ensuremath{\nu_{1}}}t}{N}}{\cos\frac{\text{\ensuremath{\nu_{1}}}t}{N}-e^{\frac{\omega_{b}}{T}}},\nonumber \\
\phi_{1} & =\arctan\frac{\sinh\frac{\omega_{b}}{T}\sin\frac{\text{\ensuremath{\nu_{1}}}t}{N}}{1-\cosh\frac{\omega_{b}}{T}\cos\frac{\text{\ensuremath{\nu_{1}}}t}{N}},\label{eq:phi012}\\
\phi_{2} & =-\arctan\frac{\left(\cosh\frac{2\omega_{b}}{T}-3\right)\sin\frac{\text{\ensuremath{\nu_{1}}}t}{N}+\cosh\frac{\omega_{b}}{T}\sin\frac{2\text{\ensuremath{\nu_{1}}}t}{N}}{\sinh\frac{\omega_{b}}{T}\left(\cos\frac{2\text{\ensuremath{\nu_{1}}}t}{N}-3\right)+\sinh\frac{2\omega_{b}}{T}\cos\frac{\text{\ensuremath{\nu_{1}}}t}{N}}.\nonumber 
\end{align}
It can be seen from Eq.\,\eqref{eq:xik} that 
\begin{equation}
\sum_{\xi}e^{\frac{-\xi N\omega_{b}}{T}}\xi^{k}\left[1-\cos\left(\nu_{0}t+\nu_{1}\xi t\right)\right]\sim1/N^{k},
\end{equation}
and we approximate the amplitude term up to $O(1/N^{2})$ in Eq.\,\eqref{eq:z1}.
Then the approximation of $Z_{1}\left(t\right)$ can be obtained as

\begin{equation}
\begin{aligned}Z_{1}\left(t\right)\simeq & 1-\left(1+e^{\frac{\omega_{b}}{T}}\right)\frac{e^{\frac{N\omega_{b}}{2T}}}{Q}\sum_{\xi}e^{\frac{-\xi N\omega_{b}}{T}}\left(a_{0}+a_{1}\xi+a_{2}\xi^{2}\right)\left[1-\cos\left(\nu_{0}t+\nu_{1}\xi t\right)\right]\\
\simeq & 1+a_{1}\coth\frac{\omega_{b}}{2T}-a_{2}\coth^{2}\frac{\omega_{b}}{2T}-\frac{\sinh\frac{\omega_{b}}{T}}{NP_{-}\left(\frac{\nu_{1}t}{N}\right)}\\
 & \times\sqrt{a_{1}^{2}+\frac{a_{2}^{2}}{N^{2}}\frac{P_{+}\left(\frac{\nu_{1}t}{N}\right)}{P_{-}\left(\frac{\nu_{1}t}{N}\right)}+\frac{2a_{1}a_{2}}{N}\sqrt{\frac{P_{+}\left(\frac{\nu_{1}t}{N}\right)}{P_{-}\left(\frac{\nu_{1}t}{N}\right)}}\cos\left(\phi_{1}-\phi_{2}\right)}\cos\left(\nu_{0}t-\phi_{3}\right),
\end{aligned}
\label{eq:z11}
\end{equation}
where 
\begin{equation}
\phi_{3}=\arctan\frac{a_{1}N\sqrt{P_{-}\left(\frac{\nu_{1}t}{N}\right)}\sin\phi_{1}+a_{2}\sqrt{P_{+}\left(\frac{\nu_{1}t}{N}\right)}\sin\phi_{2}}{a_{1}N\sqrt{P_{-}\left(\frac{\nu_{1}t}{N}\right)}\cos\phi_{1}+a_{2}\sqrt{P_{+}\left(\frac{\nu_{1}t}{N}\right)}\cos\phi_{2}}.
\end{equation}
The last $\simeq$ in Eq.\,\eqref{eq:z11} comes from the approximation
${\displaystyle \frac{e^{\left(1+\frac{N}{2}\right)\frac{\omega_{b}}{T}}}{Q}=\frac{e^{\frac{(N+1)\omega_{b}}{T}}}{e^{\frac{(N+1)\omega_{b}}{T}}-1}\simeq1}$
since $N$ is sufficiently large.

\section{the approximation of $\protect\tds$\label{sec:the-approximation-of-1}}

We start with the density matrix $\rho_{s}\left(t\right)$ for the
initial states $|\pm\rangle$,
\begin{equation}
\rho_{s}(t)=\begin{bmatrix}\rho_{00}(t) & \pm\rho_{01}(t)\\
\pm\rho_{01}^{*}(t) & \rho_{11}(t)
\end{bmatrix},
\end{equation}
and the trace distance between two initial states $|\pm\rangle$ is

\begin{equation}
\tds=|\rho_{01}(t)|^{2}.
\end{equation}
According to Eqs.\,\eqref{Bloch of final}-\eqref{eq:x2t}, the matrix
element $\rho_{01}(t)$ can be obtained up to $O(1/N^{2})$ as
\begin{equation}
\begin{aligned}\rho_{01}(t)= & \frac{e^{-i\omega_{b}t}}{2}\sum_{M=-J}^{J}\frac{e^{-\frac{M\omega_{b}}{T}}}{Q}\sum_{k,l=0}^{1}\left[\frac{1}{2}+(-1)^{k}\frac{G_{M}}{2F_{M}}\right]\left[\frac{1}{2}+(-1)^{l}\frac{G_{M+1}}{2F_{M+1}}\right]e^{i\left[(-1)^{k}F_{M}+(-1)^{l}F_{M+1}\right]t}\\
\approx & \frac{e^{\frac{N\omega_{b}}{2T}}e^{-i\omega_{b}t}}{2Q}\sum_{\xi}e^{-\frac{\xi N\omega_{b}}{T}}\sum_{h=1}^{4}(\mu_{h,0}+\mu_{h,1}\xi+\mu_{h,2}\xi^{2})e^{-i(\nu_{h,0}+\nu_{h,1}\xi)t}.
\end{aligned}
\end{equation}
The Taylor expansion is valid in the regions ${\displaystyle g<\frac{\left(3N+1\right)\Delta-2\sqrt{\Delta^{2}N(2N+1)}}{2(N+1)^{2}}}$
and ${\displaystyle g>\frac{\left(3N+1\right)\Delta+2\sqrt{\Delta^{2}N(2N+1)}}{2(N+1)^{2}}}$,
the same as that in $D_{z}\left(t\right)$. The coefficients in the
summation are listed in Table\,\ref{tab1}.

\begin{table}
\begin{tabular}{cccccccccccc}
\hline 
 & ~~~~~ &  &  & ~~~~~ & \multicolumn{3}{c}{$2gN-\Delta>0$} & ~~~~~ & \multicolumn{3}{c}{$2gN-\Delta<0$}\tabularnewline
 &  & $\nu_{h,0}$ & $\nu_{h,1}$ &  & $\mu_{h,0}$ & $\mu_{h,1}$ & $\mu_{h,2}$ &  & $\mu_{h,0}$ & $\mu_{h,1}$ & $\mu_{h,2}$\tabularnewline
\hline 
$h=1$ &  & $\frac{\nu_{1}}{2N}$ & $0$ &  & $0$ & $a_{3}$ & $-a_{3}$ &  & $\frac{a_{3}}{N}$ & $a_{3}$ & $-a_{3}$\tabularnewline
$h=2$ &  & $-\frac{\nu_{1}}{2N}$ & $0$ &  & $\frac{a_{3}}{N}$ & $a_{3}$ & $-a_{3}$ &  & $0$ & $a_{3}$ & $-a_{3}$\tabularnewline
$h=3$ &  & $\nu_{0}$ & $\nu_{1}$ &  & $1-\frac{a_{3}}{N}$ & $-2a_{3}$ & $a_{3}$ &  & $0$ & $0$ & $a_{3}$\tabularnewline
$h=4$ &  & $-\nu_{0}$ & $-\nu_{1}$ &  & $0$ & $0$ & $a_{3}$ &  & $1-\frac{a_{3}}{N}$ & $-2a_{3}$ & $a_{3}$\tabularnewline
\hline 
\end{tabular}\caption{The coefficients for the first few terms up to $O\left(1/N^{2}\right)$
in the amplitude and phase series of $\rho_{01}(t)$. Here $\nu_{0}$
and $\nu_{1}$ are the parameters of $\protect\td$, and $a_{3}=\frac{4g^{2}N^{2}}{\left(2gN-\Delta\right)^{2}}$.
It turns out that the coefficients are different for $2gN-\Delta>0$
and $2gN-\Delta<0$.}

\label{tab1}
\end{table}

The summation in $\rho_{01}(t)$ can be worked out as

\begin{equation}
\begin{aligned}\sum_{\xi}e^{-\frac{\xi N\omega_{b}}{T}}e^{-i\left(\nu_{0}+\nu_{1}\xi\right)t} & \simeq\frac{e^{-i\nu_{0}t}}{1-e^{-\frac{\omega_{b}}{T}-\frac{i\nu_{1}t}{N}}},\\
\sum_{\xi}\xi e^{-\frac{\xi N\omega_{b}}{T}}e^{-i\left(\nu_{0}+\nu_{1}\xi\right)t} & \simeq\frac{e^{\frac{\omega_{b}}{T}-\frac{it(\nu_{0}N+\nu_{1})}{N}}}{N\left(e^{\frac{\omega_{b}}{T}}-e^{-\frac{i\nu_{1}t}{N}}\right)^{2}},\\
\sum_{\xi}\xi^{2}e^{-\frac{\xi N\omega_{b}}{T}}e^{-i\left(\nu_{0}+\nu_{1}\xi\right)t} & \simeq\frac{\left(e^{\frac{\omega_{b}}{T}}+e^{-\frac{i\nu_{1}t}{N}}\right)\exp\left(-\frac{it(\nu_{0}N+\nu_{1})}{N}+\frac{(N+1)\omega_{b}}{T}-\frac{N\omega_{b}}{T}\right)}{N^{2}\left(e^{\frac{\omega_{b}}{T}}-e^{-\frac{i\nu_{1}t}{N}}\right)^{3}}.
\end{aligned}
\end{equation}
It can be seen that the matrix elements $\rho_{01}(t)$ are conjugate
for $2gN-\Delta>0$ and $2gN-\Delta<0$, implying that the $\tds$
for the two cases are the same. Then trace distance between $|+\rangle,|-\rangle$
can be obtained up to $O(1/N^{2})$ as

\begin{equation}
\begin{aligned}\tds\simeq & 2\frac{a_{3}}{N}\sqrt{\frac{P_{+}\left(\frac{\nu_{1}t}{N}\right)}{P_{-}\left(\frac{\nu_{1}t}{N}\right)}}\cos\left(\nu_{0}t+\phi_{4}\right)+\frac{a_{3}^{2}P_{+}\left(\frac{\nu_{1}t}{N}\right)}{N^{2}\left(\cosh\frac{\omega_{b}}{T}-1\right)}\\
 & +\frac{\left(\cosh\frac{\omega_{b}}{T}-1\right)\left(1-\frac{a_{3}}{N}\right)^{2}}{P_{-}\left(\frac{\nu_{1}t}{N}\right)}+\frac{2\text{\ensuremath{a_{3}}}\left(\cosh\frac{\omega_{b}}{T}-1\right)\left[e^{-\frac{\omega_{b}}{T}}-\left(1-\frac{a_{3}}{N}\right)\cos\left(\frac{\nu_{1}}{N}t\right)\right]}{NP_{-}^{2}\left(\frac{\nu_{1}t}{N}\right)},
\end{aligned}
\end{equation}
where the phase $\phi_{4}$ is
\begin{equation}
\phi_{4}=\arctan\left\{ \frac{\left(e^{-\frac{\omega_{b}}{T}}\cos\frac{\nu_{1}t}{N}+\cosh\frac{\omega_{b}}{T}\right)\left[\left(N-a_{3}\right)P_{-}\frac{\nu_{1}t}{N}+\text{\ensuremath{a_{3}e^{-\frac{\omega_{b}}{T}}}}\right]-\text{\ensuremath{a_{3}}}\left(e^{\frac{\omega_{b}}{T}}\cos\frac{\nu_{1}t}{N}+\cosh\frac{\omega_{b}}{T}\right)}{\left(e^{-\frac{\omega_{b}}{T}}\cos\frac{\nu_{1}t}{N}-\cosh\frac{\omega_{b}}{T}\right)\left[\left(N-a_{3}\right)P_{-}\frac{\nu_{1}t}{N}+\text{\ensuremath{a_{3}}}e^{-\frac{\omega_{b}}{T}}\right]+\text{\ensuremath{a_{3}}}\left(e^{\frac{\omega_{b}}{T}}\cos\frac{\nu_{1}t}{N}-\cosh\frac{\omega_{b}}{T}\right)}\tan\frac{\nu_{1}t}{2N}\right\} .
\end{equation}

\section{Trace distance for arbitrary central spin states\label{sec:Trace-distance-for}}

The trace distance between two states of the central spin at time
$t$ given arbitrary initial states $\rho_{1}(0),\,\rho_{2}(0)$ can
be obtained as

\begin{equation}
D\left(t\right)=\frac{1}{2}\sqrt{\alpha_{z}D_{z}^{2}(t)+\alpha_{x}D_{x}^{2}(t)},
\end{equation}
where the coefficients $\alpha_{x}=[x_{1}(0)-x_{2}(0)]^{2}+[y_{1}(0)-y_{2}(0)]^{2}$
and $\alpha_{z}=[z_{1}(0)-z_{2}(0)]^{2}$ are determined by the two
initial states, and the functions $D_{z}^{2}(t)$ and $D_{x}^{2}(t)$
represent the trace distance given the initial states $|0\rangle,|1\rangle$
and given the initial states $|\pm\rangle$ respectively. Then the
trace distance between two arbitrary initial states can be calculated
as: 

\begin{align}
D^{2}\left(t\right)= & \frac{1}{4}\left[\alpha_{z}D_{z}^{2}(t)+\alpha_{x}D_{x}^{2}(t)\right]\nonumber \\
= & \frac{1}{4}\left\{ \alpha_{z}\left[\overline{D_{z}}-W_{z}\left(t\right)\cos\left(\nu_{0}t-\phi_{3}\right)\right]^{2}+\alpha_{x}\left[\overline{D_{x}^{2}}+W_{x}^{2}\left(t\right)\cos\left(\nu_{0}t-\phi_{4}\right)\right]\right\} \nonumber \\
= & \frac{1}{4}\left[\alpha_{z}\overline{D_{z}}^{2}+\alpha_{x}\overline{D_{x}^{2}}+\alpha_{z}W_{z}^{2}\left(t\right)\cos^{2}\left(\nu_{0}t-\phi_{3}\right)+\alpha_{x}W_{x}^{2}\left(t\right)\cos\left(\nu_{0}t-\phi_{4}\right)-2\alpha_{z}\overline{D_{z}}W_{z}\left(t\right)\cos\left(\nu_{0}t-\phi_{3}\right)\right]\nonumber \\
\simeq & \frac{1}{4}\left[\alpha_{z}\overline{D_{z}}^{2}+\alpha_{x}\overline{D_{x}^{2}}+\alpha_{x}2\frac{a_{3}}{N}\sqrt{\frac{P_{+}\left(\nu_{cr}t\right)}{P_{-}\left(\nu_{cr}t\right)}}\cos\left(\nu_{0}t-\phi_{4}\right)-2\alpha_{z}\frac{\sinh\frac{\omega_{b}}{T}}{P_{-}\left(\nu_{cr}t\right)}\frac{a_{1}}{N}\cos\left(\nu_{0}t-\phi_{3}\right)\right]\\
= & \frac{1}{4}\left[\alpha_{z}\overline{D_{z}}^{2}+\alpha_{x}\overline{D_{x}^{2}}+\frac{\cos\left(\nu_{0}t-\phi_{5}\right)}{P_{-}\left(\nu_{cr}t\right)}\right.\nonumber \\
 & \left.\times\sqrt{4\alpha_{z}^{2}\frac{a_{1}^{2}}{N^{2}}\sinh^{2}\frac{\omega_{b}}{T}+4\alpha_{x}^{2}\frac{a_{3}^{2}}{N^{2}}P_{+}\left(\nu_{cr}t\right)P_{-}\left(\nu_{cr}t\right)-8\alpha_{x}\alpha_{z}\frac{a_{1}a_{3}}{N^{2}}\sinh\frac{\omega_{b}}{T}\sqrt{P_{+}\left(\nu_{cr}t\right)P_{-}\left(\nu_{cr}t\right)}\cos\left(\phi_{3}-\phi_{4}\right)}\right].\nonumber 
\end{align}
Here the sign $\simeq$ means that the terms with $O(1/N^{2})$ are
neglected, therefore the term $\cos^{2}\left(\nu_{0}t-\phi_{3}\right)$
can be ignored, which facilitates our calculation, and the phase $\phi_{5}$
is 
\begin{equation}
\phi_{5}=\arctan\frac{\alpha_{z}a_{1}\sinh\frac{\omega_{b}}{T}\sin\phi_{3}-\alpha_{x}a_{3}\sqrt{P_{+}\left(\nu_{cr}t\right)P_{-}\left(\nu_{cr}t\right)}\sin\phi_{4}}{\alpha_{z}a_{1}\sinh\frac{\omega_{b}}{T}\cos\phi_{3}-\alpha_{x}a_{3}\sqrt{P_{+}\left(\nu_{cr}t\right)P_{-}\left(\nu_{cr}t\right)}\cos\phi_{4}}.
\end{equation}

\twocolumngrid

\bibliographystyle{apsrev4-2}
\bibliography{References}

\end{document}